\documentclass[twocolumn]{aastex7}

\usepackage{graphicx}
\usepackage{amsmath}
\usepackage[encapsulated]{CJK}
\usepackage{ucs}
\usepackage{chngcntr}
\graphicspath{{./}{figs/}}

\newcommand{\cloudy}{\texttt{Cloudy}}
\newcommand{\cue}{\texttt{Cue}}

\newcommand{\msun}{{\rm M}_{\odot}}
\newcommand{\zsol}{{\rm Z}_{\odot}}

\newcommand{\zspec}{z_{\rm spec}}
\newcommand{\Av}{A_{\rm V}}

\newcommand{\hi}{H\,\textsc{i}}
\newcommand{\hei}{He\,\textsc{i}}
\newcommand{\heii}{He\,\textsc{ii}}
\newcommand{\ha}{H$\alpha$}
\newcommand{\hb}{H$\beta$}

\newcommand{\nii}{[N\,{\sc ii}]}
\newcommand{\oiii}{[O\,{\sc iii}]}

\newcommand{\feii}{Fe\,{\sc ii}}
\newcommand{\nev}{Ne\,{\sc v}}

\newcommand{\heiiuv}{He\,$\textsc{ii}\,\lambda$1640}
\newcommand{\heiiir}{He\,$\textsc{ii}\,\lambda$4686}
\newcommand{\heiir}{He\,$\textsc{i}\,\lambda$1.08$\mu$m}

\newcommand{\kms}{\rm km\,s^{-1}}
\newcommand{\qion}{Q_{\rm ion}}

\newcommand{\qionhe}{Q_{\rm ion, He\textsc{ii}}}
\newcommand{\vturb}{v_{\rm turb}}

\newcommand{\fagn}{f_{\rm AGN}}
\newcommand{\aion}{\alpha_{\rm ion}}
\newcommand{\bopt}{\beta_{\rm opt}}

\newcommand{\rd}{RUBIES-EGS-49140}
\newcommand{\sfit}{{\texttt{unite}}}

\shorttitle{Missing Ionizing Photons of LRDs}
\shortauthors{Wang et al}

\begin{document}
\begin{CJK*}{UTF8}{gbsn}
\title{The Missing Hard Photons of Little Red Dots: Their Incident Ionizing Spectra Resemble Massive Stars}

\correspondingauthor{Bingjie Wang}
\email{bjwang@princeton.edu}
\author[0000-0001-9269-5046]{Bingjie Wang (王冰洁)}
\thanks{NHFP Hubble Fellow}
\affiliation{Department of Astronomy \& Astrophysics, The Pennsylvania State University, University Park, PA 16802, USA}
\affiliation{Institute for Computational \& Data Sciences, The Pennsylvania State University, University Park, PA 16802, USA}
\affiliation{Institute for Gravitation and the Cosmos, The Pennsylvania State University, University Park, PA 16802, USA}
\affiliation{Department of Astrophysical Sciences, Princeton University, Princeton, NJ 08544, USA}
\email{bjwang@princeton.edu}

\author[0000-0001-6755-1315]{Joel Leja}
\affiliation{Department of Astronomy \& Astrophysics, The Pennsylvania State University, University Park, PA 16802, USA}
\affiliation{Institute for Computational \& Data Sciences, The Pennsylvania State University, University Park, PA 16802, USA}
\affiliation{Institute for Gravitation and the Cosmos, The Pennsylvania State University, University Park, PA 16802, USA}
\email{joel.leja@psu.edu}

\author[0000-0003-1561-3814]{Harley Katz}
\affiliation{Department of Astronomy \& Astrophysics, University of Chicago, Chicago, IL 60637, USA}
\email{harleykatz@uchicago.edu}

\author[0000-0001-9840-4959]{Kohei Inayoshi}
\affiliation{Kavli Institute for Astronomy and Astrophysics, Peking University, Beijing 100871, China}
\email{inayoshi@pku.edu.cn}

\author[0000-0001-7151-009X]{Nikko J. Cleri}
\affiliation{Department of Astronomy \& Astrophysics, The Pennsylvania State University, University Park, PA 16802, USA}
\affiliation{Institute for Computational \& Data Sciences, The Pennsylvania State University, University Park, PA 16802, USA}
\affiliation{Institute for Gravitation and the Cosmos, The Pennsylvania State University, University Park, PA 16802, USA}
\email{cleri@psu.edu}

\author[0000-0002-2380-9801]{Anna de Graaff}
\affiliation{Max-Planck-Institut f\"ur Astronomie, K\"onigstuhl 17, D-69117, Heidelberg, Germany}
\email{degraaff@mpia.de}

\author[0000-0002-4684-9005]{Raphael E. Hviding}
\affiliation{Max-Planck-Institut f\"ur Astronomie, K\"onigstuhl 17, D-69117, Heidelberg, Germany}
\email{hviding@mpia.de}


\author[0000-0002-8282-9888]{Pieter van Dokkum}
\affiliation{Department of Astronomy, Yale University, New Haven, CT 06511, USA}
\email{pieter.vandokkum@yale.edu}

\author[0000-0002-5612-3427]{Jenny E. Greene}
\affiliation{Department of Astrophysical Sciences, Princeton University, Princeton, NJ 08544, USA}
\email{jgreene@astro.princeton.edu}

\author[0000-0002-2057-5376]{Ivo Labb\'e}
\affiliation{Centre for Astrophysics and Supercomputing, Swinburne University of Technology, Melbourne, VIC 3122, Australia}
\email{ilabbe@swin.edu.au}

\author[0000-0003-2871-127X]{Jorryt Matthee}
\affiliation{Institute of Science and Technology Austria, Am Campus 1, Klosterneuburg, Austria}
\email{jorryt.matthee@ist.ac.at}

\author[0000-0002-2446-8770]{Ian McConachie}
\affiliation{Department of Astronomy, University of Wisconsin-Madison, Madison, WI 53706, USA}
\email{ian.mcconachie@wisc.edu}

\author[0000-0003-3729-1684]{Rohan P. Naidu}
\thanks{NHFP Hubble Fellow}
\affiliation{MIT Kavli Institute for Astrophysics and Space Research, Cambridge, MA 02139, USA}
\email{rnaidu@mit.edu}

\author[0000-0002-7524-374X]{Erica J. Nelson}
\affiliation{Department of Astrophysical and Planetary Science, University of Colorado, Boulder, CO 80309, USA}
\email{erica.june.nelson@colorado.edu}

\begin{abstract}

The nature of Little Red Dots (LRDs) has largely been investigated through their continuum emission, with lines assumed to arise from a broad-line region. In this paper, we instead use recombination lines to infer the intrinsic properties of the central engine. Our analysis first reveals a tension between the ionizing properties implied from H$\alpha$ and He\,\textsc{ii}$\,\lambda$4686. The high H$\alpha$ EWs require copious H-ionizing photons, more than the bluest AGN ionizing spectra can provide. In contrast, He\,\textsc{ii} emission is marginally detected, and its low EW is, at most, consistent with the softest AGN spectra. The low He\,\textsc{ii}/H$\beta$ ($\sim10^{-2}$, $<20\times$ local AGN median) further points to an unusually soft ionizing spectrum. We extend our analysis to dense gas envelopes (``quasi-star''/``black-hole star''), and find that hydrogen recombination lines become optically thick and lose diagnostic power, but He\,\textsc{ii} remains optically thin and a robust tracer. Photoionization modeling with \texttt{Cloudy} rules out standard AGN accretion disk spectra. Alternative explanations include: exotic AGN with red rest-optical emission; high {\it{average}} optical depth ($>10$) from gas/dust; and/or soft ionizing spectra with abundant H-ionizing photons, consistent with e.g., a cold accretion disk or a composite of AGN and stars. The latter is an intriguing scenario since high hydrogen densities are highly conducive for star formation, and nuclear star clusters are found in the vicinity of local massive black holes. While previous studies have mostly focused on features dominated by the absorbing hydrogen cloud, the He\,\textsc{ii}-based diagnostic proposed here represents a crucial step toward understanding the central engine of LRDs.

\end{abstract}


\section{Introduction}

The discovery of a new class of compact, red objects in JWST deep fields---commonly referred to as Little Red Dots (LRDs; \citealt{Matthee2024})---has presented a major puzzle.
Characterized by a compact morphology, a red rest-frame optical continuum, a faint ultraviolet (UV) component, and broad Balmer emission lines (e.g., \citealt{Furtak2024,Greene2024,Wang2024:ub,Labbe2024:monster,Kocevski2025}), LRDs exhibit a unique combination of properties that challenges conventional models of active galactic nuclei (AGN) and their host galaxies. Their nature remains elusive despite intense scrutiny.

Early after their discovery, the pronounced spectral break near the Balmer limit in the luminous LRDs motivated joint AGN and host galaxy modeling where the UV and Balmer break are attributed to stars, and an AGN dominates the rest optical light \citep{Wang2024:ub,Wang2024:brd,Labbe2024:monster,Ma2025}.
Such modeling appeared to be well-motivated because a break around the Balmer limit is traditionally associated with evolved stellar populations, emerging after star formation has declined significantly for at least $\sim 100$~Myr \citep{Bruzual1983,Hamilton1985,Worthey1994,Balogh1999}. 
This interpretation implies an extremely rapid and early episode of star formation, potentially in tension with theoretical limits \citep{Boylan-Kolchin2023,Labbe2023:ub,Wang2024:ub}.

More recently, \citet{Inayoshi2025:dense} proposed a non-stellar origin for the Balmer break, attributing it to absorption by dense neutral hydrogen gas that surrounds an AGN.
This mechanism becomes increasingly appealing in light of recent reports of LRDs displaying extremely strong Balmer breaks---stronger than any predicted by standard stellar population models \citep{deGraaff2025:cliff,Naidu2025}.

An AGN interpretation, made viable under the dense gas envelope hypothesis (often referred to as a ``quasi-star''; \citealt{Begelman2008} or a ``black-hole star"; \citealt{Naidu2025}), offers a natural explanation for the presence of broad Balmer emission lines. However, key observational signatures expected for standard AGNs remain absent or contradictory. Although the hypothesis of accreting black holes inside gas envelopes provides a potential pathway to explain those contradictions, it is nevertheless instructive to consider the seemingly conflicting observational evidence.
Notably, flat MIRI detections (\citealt{Wang2024:brd,deGraaff2025:cliff}; see also \citealt{Perez-Gonzalez2024,Williams2024,Akins2024}), probing the rest-frame mid-infrared, and non-detections of dust continuum emission in ALMA observations \citep{Akins2025:alma,Casey2025,Setton2025:alma,Xiao2025} suggest a lack of hot dusty torii, a feature nearly ubiquitous in AGNs.
A lack of X-ray detections poses another challenge to an AGN interpretation: individual sources remain undetected \citep{Labbe2024:monster,Wang2024:brd,Deugenio2025:absorption}, and stacked analyses yield, at most, marginal signals \citep{Ananna2024,Yue2024,Sacchi2025}.
The general lack of variability adds to the ambiguity \citep{Zhang2025:var}, though tentative variability has been reported in a few cases \citep{Furtak2025,Ji2025,Naidu2025}. If confirmed, however, variability would be a strong piece of evidence against a stellar origin, as normal galaxies do not exhibit variability unless a supernova happens to occur within a narrow time window of observation. Interestingly, the weakness in both X-rays and variability can be explained by super-Eddington accretion \citep{Inayoshi2024:supedd}.
Finally, the origin of the extended UV emission remains an open question. The resolved UV morphologies measured in bright LRDs \citep{Wang2024:brd,Baggen2024} from the RUBIES program \citep{deGraaff2024:survey}, as well as in other independent studies \citep{Chen2025:host,Rinaldi2025}, are difficult to reconcile with a simple AGN model. Proposed explanations include UV emission from the host galaxy (e.g., \citealt{Wang2024:brd}), or scattered AGN light (e.g., \citealt{Greene2024}).

Until now, the main focus has been on the source of the continuum of LRDs, while the emission lines have been treated as emerging from a broad-line region ionized by a standard AGN accretion disk. These lines have been used to infer surprisingly high black hole masses following typical reverberation mapping scaling relationships (e.g., \citealt{Greene2005}). Here, we interrogate those assumptions directly by probing the ionizing properties of LRDs, and inferring the intrinsic properties of the central engine.
We use recombination lines for this task. Their constraints on the ionizing mechanisms of LRDs are examined by comparing the predicted emission-line spectra with observations.
Under ionization-bounded photoionization conditions typical in AGN, the rate of recombination, and therefore, the strength of recombination lines, is determined primarily by the ionizing photon budget. This means that, in principle, it would be possible to infer the intrinsic shape and normalization of the ionizing spectrum by studying multiple recombination lines.

Our approach is consistent with the method first proposed in \cite{Zanstra1929}, but also bears some resemblance to the ``softness parameter'', defined by line ratios from different ions of the same element \citep{Vilchez1988}. The later has been used to constrain the hardness of the ionizing radiation in H\textsc{ii} or star forming regions (e.g., \citealt{Simon-Diaz2008}), with recent extensions to AGN SEDs (e.g., \citealt{Perez-Montero2025}. However, its direct application to LRDs is limited by the available spectral coverage and the lack of required key emission-line ratios.

Crucially, the dense gas envelope hypothesis raises the possibility that the intrinsic properties of the central engine in LRDs may be fundamentally inaccessible.
The neutral hydrogen must reach extremely high column densities and number densities, so that the effective optical depth for hydrogen blue-ward of the Balmer limit becomes much greater than 1 to produce a Balmer break.
For example, these values reach $N_{\rm H} =10^{24}~{\rm{cm^{-2}}}$, $n_{\rm H} =10^{10}~{\rm{cm^{-3}}}$ for A2744-QSO1 in \citet{Ji2025}, and $N_{\rm H} =10^{25.8}~{\rm{cm^{-2}}}$, $n_{\rm H} =10^{11}~{\rm{cm^{-3}}}$ for MoM-BH*-1 in \citet{Naidu2025}.
At the limit of very high optical depths and densities, such objects should emit effectively as blackbodies with single temperatures; in this scenario, most or all information about the central power source is fundamentally inaccessible, analogous to the case of a stellar photosphere in, e.g., the black hole envelope model of \citet{Kido2025}.

In this paper, we consider both the standard case, in which no dense gas is present, and scenarios involving dense neutral hydrogen, to explore the limits and robustness of our proposed recombination-line diagnostics.
The structure of this paper is as follows.
Section~\ref{sec:data} introduces the dataset.
Section~\ref{sec:measure} details measurements of recombination lines.
Section~\ref{sec:toy} presents a toy model to establish the relationship between the intrinsic spectral shape and recombination line fluxes.
Section~\ref{sec:cloudy} outlines the setup of our photoionization modeling.
Section~\ref{sec:res} presents the constraints on the LRD intrinsic spectra.
Section~\ref{sec:dis} discuss the implications of our findings for the nature of LRD, considering both standard and dense gas scenarios.
We conclude in Section~\ref{sec:concl}.

Where applicable, we adopt the best-fit cosmological parameters from the WMAP 9 yr results: $H_{0}=69.32$ ${\rm km \,s^{-1} \,Mpc^{-1}}$, $\Omega_{M}=0.2865$, and $\Omega_{\Lambda}=0.7135$ \citep{Hinshaw2013}, and the stellar initial mass function from \citet{Chabrier2003}.

\section{Data\label{sec:data}}

\subsection{RUBIES LRD Sample}

\begin{figure} 
\gridline{
  \fig{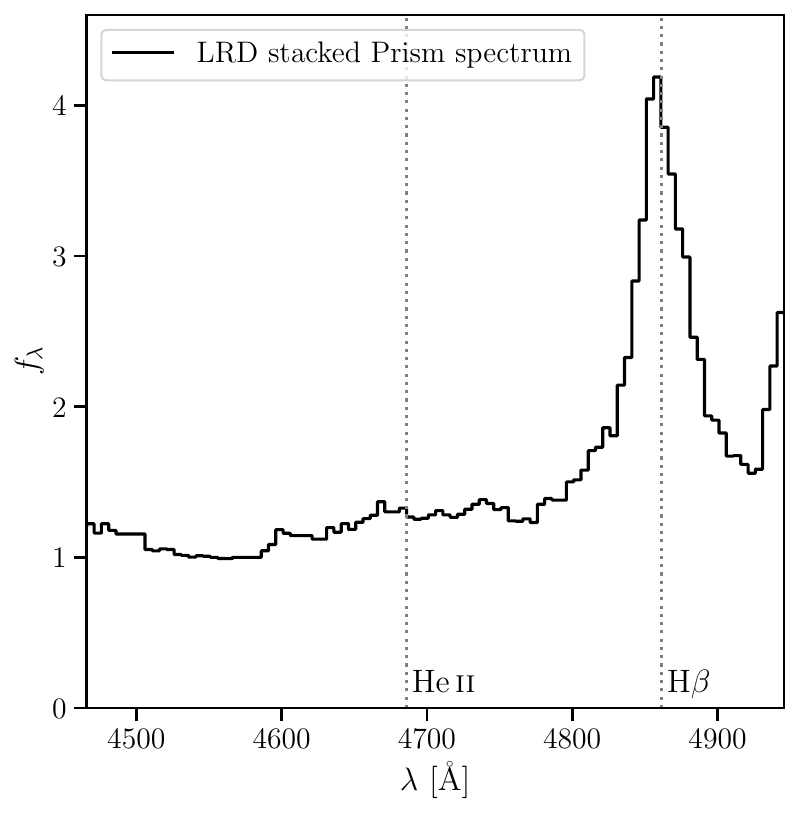}{0.45\textwidth}{}
}
\caption{Stacked Prism spectrum of LRDs, zoomed in to the spectral region around the \heii\ and \hb\ lines. The stack is from a uniformly selected sample of LRDs from RUBIES \citep{Hviding2025}, providing complementary information for LRDs as a population.
}
\label{fig:stack}
\end{figure}

The high spectroscopic completeness for red objects and the homogeneous data quality of the RUBIES program (JWST-GO-4233; PIs de Graaff \& Brammer; \citealt{deGraaff2024:survey}) enables the construction of a uniformly selected spectroscopic sample of LRDs \citep{Hviding2025}.
This sample is selected based on three criteria: the presence of broad Balmer emission lines, v-shaped continua, and a dominant point source in the rest-frame optical.
We use this sample primarily for illustrative purposes, demonstrating that our results are representative of the LRD population on average.

The RUBIES program is designed with an exposure time of 48 minutes for both the Prism/Clear and the G395M/F290LP modes. All spectra are reduced with \texttt{msaexp} \citep{Brammer2022}, corresponding to version 4 on the DAWN JWST Archive\footnote{\url{https://dawn-cph.github.io/dja}}.
We construct stacked spectra using three methods: simple mean, median, and signal-to-noise-weighted average. Since the intent is merely to confirm the weakness of the \heii\ emission, we do not apply a detailed treatment of the wavelength-dependent line spread function (LSF). Instead, we shift each individual spectrum to the rest frame and interpolate them onto a common wavelength grid while conserving fluxes.
A zoom-in near \hb\ of the median stacked spectrum is shown in Figure~\ref{fig:stack}.

Several well-studied LRDs are included in this sample: RUBIES-EGS-49140, 55604, and 966323 at $z \gtrsim 7$ from \citet{Wang2024:ub}, which exhibit pronounced Balmer breaks and were initially selected as massive galaxy candidates \citep{Labbe2023:ub};
RUBIES-BLAGN-1 at $z = 3.1$, notable for its exceptional brightness and surprisingly flat MIRI SED, suggesting a lack of a hot dusty torus \citep{Wang2024:brd};
and ``The Cliff'' \citep{deGraaff2025:cliff}, which has one of the strongest Balmer breaks detected to date.
A set of G395M spectra are supplemented in Appendix~\ref{app:sample}.
While the S/N of most of the grating spectra ($\sim 0 - 4$, with a median of 1) are insufficient for robust measurements of \heii\ EWs, they demonstrate that the low \heii\ EWs found in LRDs in general are not an effect of the low resolution of Prism spectra.

In addition, considering the diverse range of spectral features observed in LRDs, and the ongoing efforts toward a physical model to explain their nature, we further assemble a sample of LRDs drawn from various independent studies.
Each selected object has been previously studied in detail, which is a choice made deliberately to let the existing analyses to provide valuable context for comparison and interpretation.
This supplementary sample serves to ensure the general applicability of our results. Details of these sources are also provided in Appendix~\ref{app:sample}.

\subsection{RUBIES-EGS-49140}

\begin{figure*} 
\gridline{
  \fig{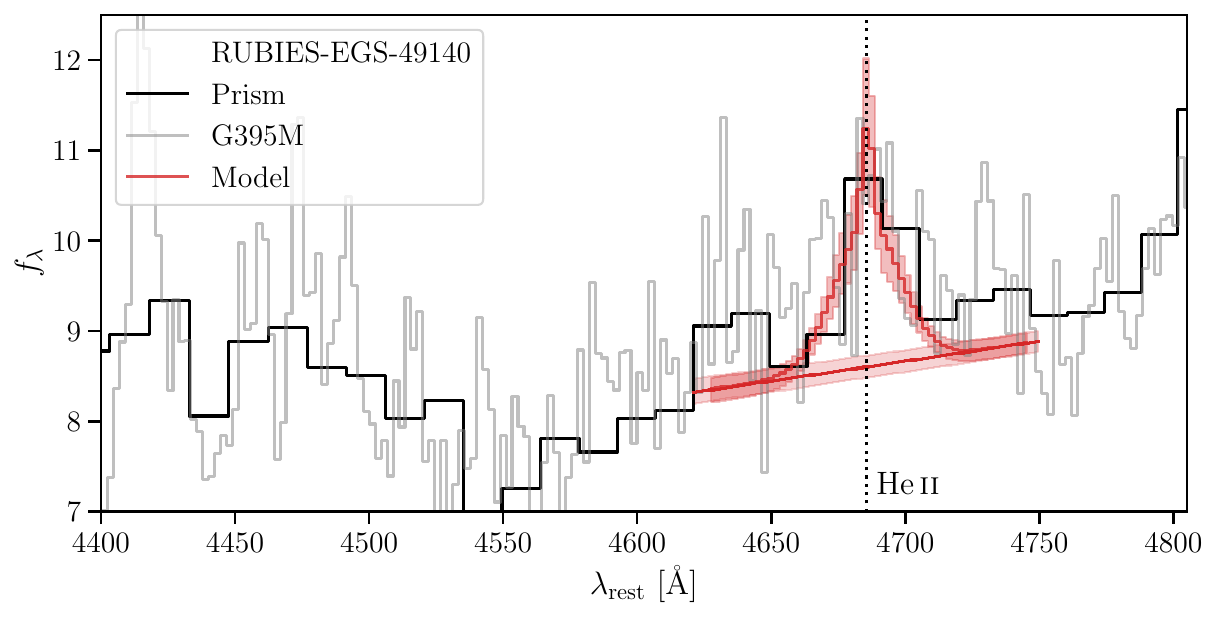}{0.9\textwidth}{}
}
\caption{Zoom-in around the \heii\ region. The Prism and the medium-resolution spectra of \rd\ from PID 4106 are shown in black and gray, respectively. Modeled line profile (a narrow + a broad Gaussian component) and the local continuum used are over-plotted in red. The uncertainties in the measured \heii\ EW may arise from the unknown kinematics as well as ambiguity in the local continuum. Nevertheless, it is evident that \heii\ is weak, even in an 8~hr G395M spectrum.}
\label{fig:data}
\end{figure*}

\rd\ is a prototypical LRD, first identified spectroscopically in \citet{Wang2024:ub} as part of the RUBIES program \citep{deGraaff2024:survey}. In April 2025, \rd\ was reobserved in JWST-GO-4106 (PIs: Nelson \& Labb\'e), yielding significantly deeper spectroscopy: 2.9 hours with the PRISM and 7.7 hours with the G395M grating \citep{DEugenio2025:fe}. Owing to the improved signal-to-noise ratio (S/N), we focus on these new deep spectra in our analysis. Zoom-in near \heii\ is shown in Figure~\ref{fig:data}.

\subsection{Local AGNs\label{sec:data:sdss}}

\begin{figure*} 
\gridline{
  \fig{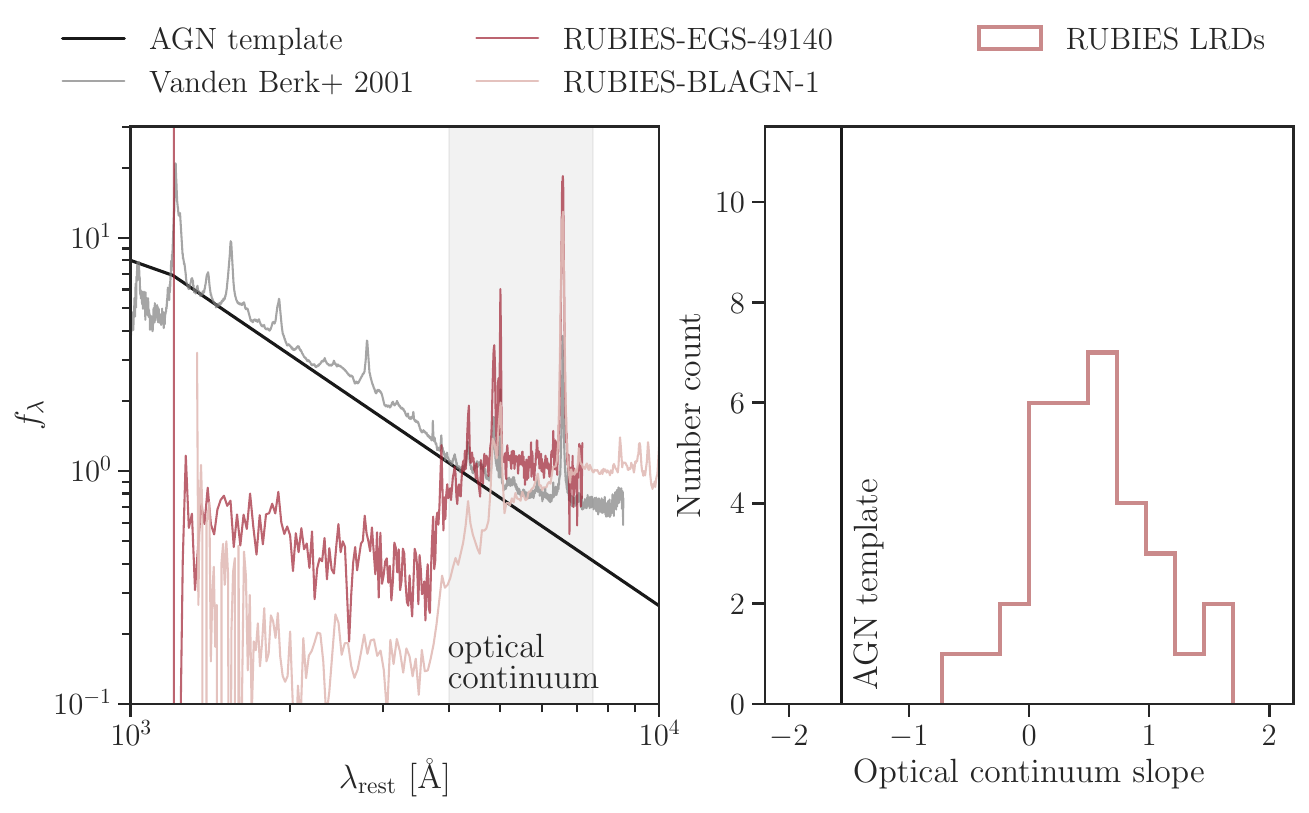}{0.9\textwidth}{}
}
\caption{Comparison between the spectra of typical AGNs and LRDs. (Left) A broken power-law fit (black) to the SDSS quasar composite spectrum (gray) is taken to be the fiducial AGN template \citep{VandenBerk2001}. Two example LRD spectra from \citet{Wang2024:ub, Wang2024:brd} are shown in red for comparison.
(Right) The histogram shows the distribution of optical continuum slopes measured for the RUBIES LRD sample \citep{Hviding2025}, with the typical AGN slope indicated by the vertical black line. The UV-optical continuum from an AGN accretion disk corresponds to the Rayleigh-Jeans tail, meaning that the expected variation in the slope is small across blackbody temperatures.
It is evident that the observed LRD spectra are significantly redder than typical AGNs.
}
\label{fig:prism}
\end{figure*}

We take the empirical composite quasar spectrum from the Sloan Digital Sky Survey (SDSS) presented in \citet{VandenBerk2001} to serve as a typical AGN spectrum for comparison. We adopt their broken power-law best-fit to the observed UV-optical AGN continuum.
As shown in Figure~\ref{fig:prism}, the continuum slope changes abruptly near 5000~\AA. \citet{VandenBerk2001} attribute this change to contamination from host galaxies. 
Given the lack of evidence of substantial host galaxy contributions in the rest optical/IR range of the LRD spectra, here we opt to only consider the emission from the accretion disk as the dominant source of flux across the full wavelength range. We therefore neglect the change in the slope and adopt a single power-law to describe the optical AGN continuum, consistent with e.g., \citet{Glikman2006,Temple2021}.
Physically, the single power law is also consistent with the expectation that the optical continuum from an AGN accretion disk corresponds to the Rayleigh-Jeans tail, which generally exhibits minimal variation across blackbody temperatures.

In addition, we utilize SDSS AGN as reference samples. The \ha\ EWs are taken from the SDSS quasar catalog, compiled as part of the 16th data release \citep{Paris2018}.
These targets are classified into stars, galaxies and quasars by fitting spectra using stellar templates and a principal component analysis decomposition of galaxy and quasar spectra.

Since emission-line diagnostics are commonly used for separating AGNs from star-forming galaxies, we also examine the \heii-based diagnostic proposed by \citet{Shirazi2012}.
In particular, \citet{Bar2017} applied this diagnostic to identify $\sim 1300$ AGNs in SDSS via the \heii/\ha\ versus \nii/\hb\ diagram.
Their sample is smaller than the \citet{Paris2018} catalog, primarily due to selection criteria: 
First, they limit the redshift range to $0.02 < z < 0.05$ and include only the galaxies with absolute Petrosian r-band magnitude $<-19.0$ to obtain a magnitude-limited sample.
Second, they further require all four emission lines used in the diagnostic to have S/N $>3$.
In our analysis, we follow \citet{Bar2017} by adopting their AGN sample and using nebular emission line measurements from the OSSY catalog \citep{Oh2011}.

\section{Measurements\label{sec:measure}}

\subsection{Hydrogen and Helium Recombination Lines\label{sec:line}}

The emission lines of interest (the \ha\ and \nii\ complex, \hb, \oiii, and \heii) in the Prism and G395M spectra are fit simultaneously using the \sfit\ package, the setup of which largely follows that of \citet{Hviding2025}.
Here we only reiterate the key modeling components.

\sfit\ accounts for the under-sampling of the line spread function (LSF) in NIRSpec by integrating the model in each pixel. We use the NIRSpec LSF curves assuming an idealized point source generated with \texttt{msafit} \citep{deGraaff2024:msafit}. In addition, \sfit\ has a nuisance parameter to broaden the LSF to capture its systematic uncertainty. The Balmer emission lines are modeled with four Gaussians: a narrow emission, a broad emission, a second broad emission, and an absorption. For the \heii\ line, we try two models due to the uncertain kinematics: one model is described by a narrow Gaussian only, whereas a second model is described by a narrow and a broad Gaussian.
As the physical origin of the emission lines remains a major open question, we consider both possibilities plausible. Together, these two models are intended to bracket the uncertainty in the \heii\ line flux.
The \oiii\ and \nii\ flux ratios are fixed to the quantum ratios of 1:2.98 and 1:2.95, respectively \citep{Galavis1997}.
All narrow emission line models share a common redshift and intrinsic velocity width, and the same is for all the broad emission line models.

Similar to the findings for \ha\ in \citealt{Hviding2025}), the two-component narrow+broad Gaussian model underestimates the extended wings of the \hb\ emission (see Appendix~\ref{app:linefit}).
We thus add a second broad Gaussian component to capture the flux in the wings.
While this paper does not focus on kinematics, the \hb\ line is used solely in tying the different Gaussian components during our fitting procedure, and later in computing the \heii/\hb\ flux ratios.
It is worth emphasizing that none of the results depend on the $\sim~10$\% shifts in \hb\ flux that arise when including or excluding this second broad Gaussian component.
\citet{Rusakov2025} recently performed detailed line profile fitting for a sample of 12 LRDs. When comparing to the \ha\ EWs of SDSS AGNs (Figure~\ref{fig:toy}), we adopt their \ha\ EWs. However, we confirm that our own measurements are consistent with their values.
Furthermore, we note that different scattering mechanisms---resonant, Raman, and Thomson scattering---produce distinct line profiles. These effects have been studied in detail in recent works (e.g., \citealt{Chang2025, Kokorev2025, Torralba2025}). In this paper, we are concerned only with the integrated line flux. For this purpose, our modeling approach is sufficient.

\subsection{Rest Optical Continuum Shape\label{subsec:bopt}}

The red rest-optical continuum is one of the defining characteristics of LRDs, and has motivated theoretical models invoking, for instance, super-Eddington accretion \citep{Lambrides2024,Liu2025}. Dense gas may also introduce changes to the optical continuum shape \citep{deGraaff2025:cliff,Naidu2025}. We thus additionally quantify the optical continuum slope, $\bopt$, which will be used later to constrain the properties of the central engine in our analysis.

The shape of the rest-frame optical continuum is characterized by fitting a power law to the observed spectra. The fit is performed over the wavelength range between 4000 and 7500~\AA, indicated in gray shading in Figure~\ref{fig:prism}, and all strong emission lines are masked.
We note that this definition is different from \citet{Hviding2025}, where the wavelength range includes the Balmer break region.

It is worth pointing out that such a simple power-law fit does not necessarily capture the shape of the optical continua of LRDs, particularly in cases where strong spectral breaks are present. We only use the optical slopes as a first-order approximation for the observed redness of LRDs.
This approach is intended to be illustrative, serving to highlight the contrast with the much bluer optical continua typically expected from standard AGNs, rather than to provide a detailed physical model of the continuum shape.

\section{Demonstrating the Connection between Recombination Lines and the Intrinsic Ionizing Spectrum\label{sec:toy}}

As a demonstration that the intrinsic properties of the ionizing spectrum can, in principle, be inferred from recombination lines, we construct a toy model.

Among the hydrogen recombination lines, we focus on \ha, as it is typically one of the strongest and most commonly observed features in LRD spectra.
In addition, we consider \heiiir\ emission. The ionization potential of He$^{++}$ is 54.4 eV, which is $4\times$ higher than that of hydrogen. This makes it a useful indicator for the presence of hard ionizing radiation that is usually not expected from normal star-forming galaxies. The consideration of \heii\ as a probe of the ionizing continuum has a long history (e.g., \citealt{Penston1978,Ferland1983,Pequignot1984, Kehrig2018, Mondal2025, Roy2025}) and has been incorporated into revised emission-line diagnostics designed to better distinguish AGNs from star-forming systems (e.g., \citealt{Shirazi2012}).

\begin{figure*} 
\gridline{
  \fig{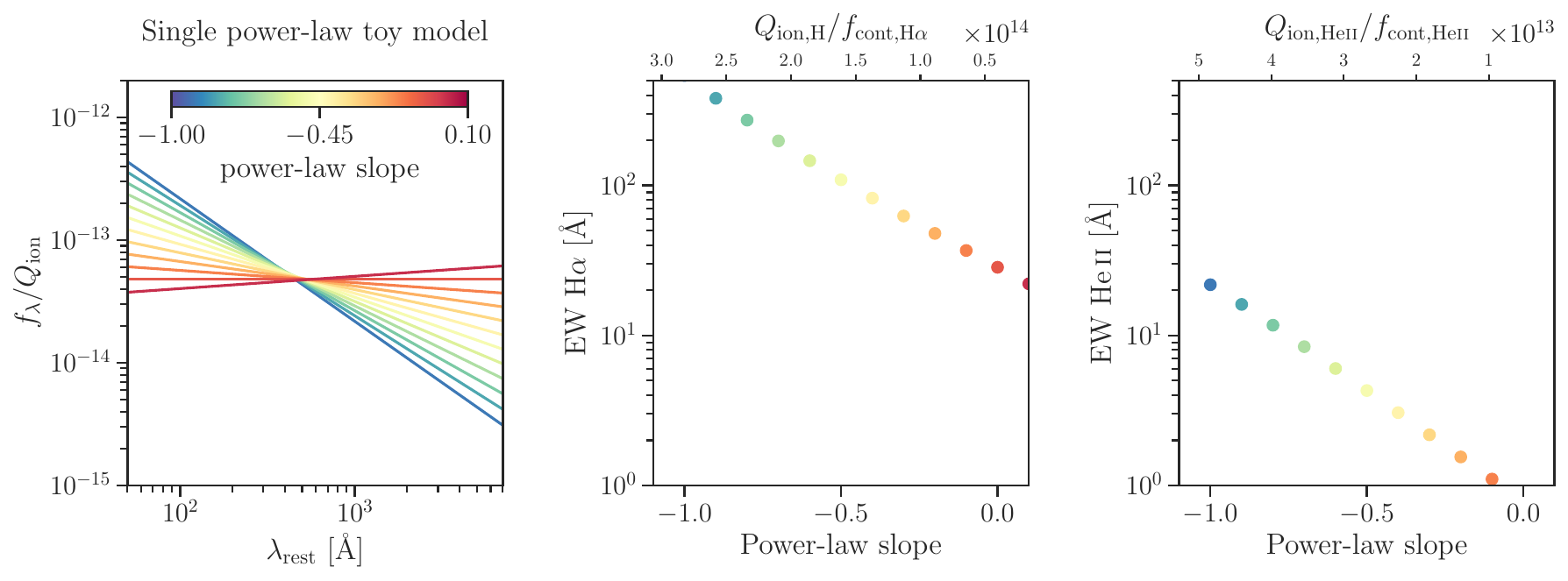}{0.9\textwidth}{(a)}
} 
\gridline{
  \fig{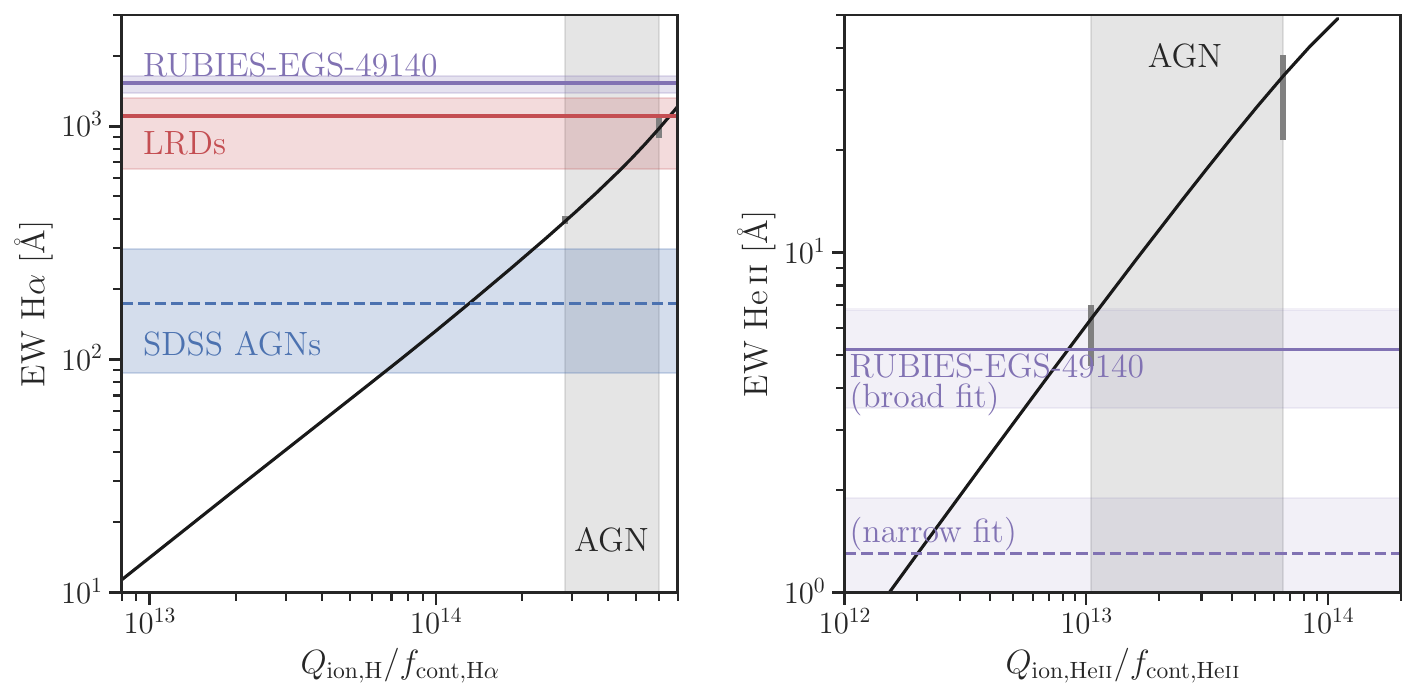}{0.9\textwidth}{(b)}
}
\caption{Relationship between intrinsic ionizing spectrum and recombination lines.
(a) The predicted \ha\ and \heii\ EWs are plotted as functions of power-law slopes when the input SEDs are all described by single power laws. 
Example spectra, normalized by the number of hydrogen-ionizing photons, $\qion$, are plotted in the first panel in the same color codes.
A large line EW corresponds to a steeper power law.
(b) Similar to the upper panels, but using incident SEDs that more realistically represent AGNs; i.e., the AGN template shown in Figure~\ref{fig:prism}. 
The gray shaded region marks the expected range of the AGN ionizing spectra, whereas the black lines represent model predictions assuming typical elemental abundances.
The median of the \ha\ EWs of LRDs is plotted as a red horizontal line, with red shading indicating the 16th and 84th quantiles. For comparison, the distribution of \ha\ EWs of SDSS AGNs is plotted in the same format in blue.
\rd\ is highlighted in purple.
For the \heii\ EW, we quote two measured values for \rd\ to reflect the potential systematic uncertainties: with and without assuming a broad Gaussian component.
Regardless of the modeling choice, a clear inconsistency emerges: the large \ha\ EW observed in \rd\ implies an ionizing spectrum that is harder than that of typical AGNs, whereas the low \heii\ EW suggests a much softer ionizing spectrum.}
\label{fig:toy}
\end{figure*}

\subsection{Single Power Law\label{subsec:pl}}

To gain intuition for how the ionizing spectrum affects recombination lines, we begin with a simple model in which the full SED is described by a single power law, $f_\lambda \propto \lambda^{\alpha}$. We vary the power-law index $\alpha$ to explore a range of spectral slopes.
The typical range for AGN spans $-0.8$ to $0$ (e.g., \citealt{Zheng1997,Groves2006,Lusso2015,Feltre2016}).

These model spectra are used as inputs to a \cloudy\ emulator, \cue\ \citep{Li2024}, to obtain the line fluxes. Examples are included in Figure~\ref{fig:toy}.
This works because, for ionized gas in an ionization-bounded nebulae, \ha\ flux is insensitive to the shape of the ionizing spectrum or other nebular parameters (e.g., density, temperature) under typical conditions of the interstellar medium, and is instead very tightly correlated with the number of ionizing photons, $Q_{\rm ion}$.

No dust is included, and the stopping criteria are set by either the kinetic temperature dropping below 100 K or the ratio of electron to total hydrogen number densities falling below 0.1. A spherical geometry with a unity filling factor is assumed. While some studies adopt plane-parallel models (e.g., \citealt{Thomas2018, Perez-Diaz2022}), in most cases the choice of geometry affects the predicted quantities at only the 10-15\% level \citep{Ferland2006}.

As for the rest of the \cue\ parameters including the ionization parameter, hydrogen density, and elemental abundances, we simply use the median of the priors for this simple model for illustration. Variations in the line fluxes due to these parameters are estimated by drawing random sample over the full uniform prior ranges used in \citet{Li2024}:
log~$U ~\in$ [-4, 1], log~$n_{\rm {H}}/{\rm cm^{-3}}~\in$ [1, 4], log~(O/H)/(O/H)$_\odot$ $\in$ [-2.2, 0.5], log~(C/O)/(C/O)$_\odot$ $\in$ [-1.0, 0.7], and log~(N/O)/(N/O)$_\odot$ $\in$ [-1.0, 0.7].
The typical ranges for \ha\ and \heii\ line fluxes are about 0.2 dex.

\subsection{AGN\label{subsec:toy_agn}}

The SEDs of AGNs can vary and are known to exhibit complex dependencies on parameters such as viewing angle. A detailed treatment of these effects is beyond the scope of this work. For simplicity and interpretability, we adopt a broken power-law best-fit to the composite SDSS quasar spectrum \citep{VandenBerk2001}, shown as black lines in the left panel of Figure~\ref{fig:prism}.

Since the SDSS composite does not extend into the wavelength range of the ionizing spectrum, we extrapolate the spectrum short-ward of 1200~\AA\ by appending a power-law. We then vary the power-law slope in this ionizing portion of the SED and use the resulting spectrum as input to \cue\ to predict the expected emission line fluxes. 

We parameterize this slope by the number of hydrogen (helium) ionizing photons normalized by the continuum flux near \ha\ (\heii).
For illustrative purposes, Figure~\ref{fig:toy}-a shows how this parameterization corresponds to single power-law slopes.
Detailed discussions of the results shown in Figure~\ref{fig:toy} are provided in Sections~\ref{sec:res:toy}, \ref{sec:res:haheii}, and \ref{sec:dis:turnover}.

\section{Photoionization Modeling with Dense Hydrogen Gas\label{sec:cloudy}}

In light of the recent works explaining the observed spectral break in LRDs with gas attenuation from a dense hydrogen cloud \citep{Inayoshi2025:dense}, we perform photoionization modeling using the C23 version of \cloudy\ \citep{Chatzikos2023}.
The general setup follows that of \citet{Inayoshi2025:dense,Ji2025}. 
Specifically, we use a custom SED as the incident spectrum, and create a grid with the following varying parameters:
number density of neutral hydrogen $\log(n_{\rm H}/{\rm cm}^{-3}) = 8$ to $12$ in steps of $0.5$;
column density of neutral hydrogen $\log(N_{\rm H}/{\rm cm}^{-2}) = 20$ to $27$ in steps of $0.5$;
ionization parameter $\log U = -2.5$ to $-1$, in steps of 0.25;
and metallicity $\log (Z/\zsol) = -2.5$ to $-0.5$, in steps of 0.5.
No dust is included, and the calculation stops when the column density is reached.

An insight from \citet{Ji2025} is that introducing turbulence results in a smoother Balmer break---more consistent with the observed LRD spectra---compared to the sharp discontinuities seen in the models of \citet{Inayoshi2025:dense}.
However, we note that whether turbulence is indeed driving the smooth breaks remains an open question.
We adopt two fixed turbulence values of $\vturb = 120~\kms$ \citep{Ji2025} and $\vturb = 500~\kms$ \citep{deGraaff2025:cliff,Naidu2025}, which serve to demonstrate that the main results of this paper remain unaffected with different turbulent velocities (Appendix~\ref{app:vturb}).

The net transmitted spectra from \cloudy\ include the attenuated incident and diffuse continua and lines, and corresponds to what would be observed if the continuum source were viewed through the gas in a plane-parallel geometry.
The total spectrum includes both the transmitted and reflected components, approximating what would be observed if the source were surrounded by a spherical distribution of gas.
In the aforementioned studies, the net transmitted spectra are found to provide a better fit to the data than the total spectra. 
We thus adopt the net transmitted spectra throughout this paper.
This is a critical assumption, since the complete lack of reflected emission implies that it is either entirely absorbed or the covering fraction is close to 1.
A detailed investigation of geometric effects is beyond the scope of this paper. However, we note that including the reflected component generally increases the line fluxes and potentially the EWs, though at the expense of a weaker Balmer break. Consequently, relying on the net transmitted spectra to interpret the weak \heii\ emission can be seen as a conservative choice, as the effects from including the reflected component would only strengthen our conclusions.

\subsection{AGN Spectra}

We give special attention to the construction of the incident SEDs in the \cloudy\ modeling. 
For AGNs, we adopt the two-piece power-law model as in Section~\ref{sec:toy}.
Essentially, we take a standard AGN accretion disk as our baseline model.

We sample three values of the ionizing spectral slope in $f_\nu$ vs $\nu$: $\aion = -2, -1.6, -1.2$ (or $-0.8$ to $0$ in $f_\lambda$ vs $\lambda$), which bracket the typical range reported in the literature (e.g., \citealt{Zheng1997,Groves2006,Lusso2015,Feltre2016}).

The UV-optical continuum from an AGN accretion disk corresponds to the Rayleigh-Jeans tail, which generally exhibits minimal variation across blackbody temperatures. We take the best-fit optical slope from the \citet{VandenBerk2001} template, $\bopt=0.46$ in $f_\nu$ vs $\nu$, as the fiducial value.
We also explore redder rest-optical slopes, $\bopt = 1.2$ and $2.0$. This is an empirical choice, and only meant to be illustrative.

In addition, we adopt the empirical SED templates constructed from a sample of local, unobscured Type~1 AGNs \citep{Jin2012}. These templates represent median SEDs binned by Eddington ratios; varying the accretion rate modifies both the slope of the ionizing spectrum and the shape of the optical continuum. As such, they provide a complementary perspective.

\subsection{AGN and Stellar Composite Spectra\label{subsec:toy_star}}

We also consider composite spectra of varying degrees of contributions from stars and AGNs.
For the stellar component, we adopt a 10~Myr-old, metal-poor simple stellar population (SSP) with 10\% solar metallicity, based on the MIST stellar isochrones \citep{Choi2016,Dotter2016} and MILES stellar library \citep{Sanchez-Blazquez2006} from FSPS \citep{Conroy2010}.
While this choice reflects the expectation for young, chemically unevolved systems in the early universe, one might expect that the metallicity of the stellar component to be the same as that of gas. As an additional test, we generate SSPs with metallicities identical to those assumed for the gas, and perform the same analyses.

The specific choice of the stellar population synthesis (SPS) model is not central to our analysis and is intended to be illustrative. A variety of SPS models exist, each incorporating different assumptions (e.g., \citealt{Leitherer1999, Bruzual2003, Eldridge2017}). Depending on stellar age and gas conditions, the inclusion of, for example, binary stellar populations can alter the predicted line emissions (e.g., \citealt{Perez-Diaz2022, Gonzalez-Diaz2025}).
However, we note that the MIST stellar isochrones include rotating stars, the \heii\ outputs of which have been shown to be similar to those of binaries \cite{Choi2017}.
In any case, since our goal is to explore the effect on inference of the central engine in LRDs in the presence of a stellar-like contribution, we consider an exploration based on one SSP model to be sufficient for the purposes of this work.

The fractional contribution of stars vs. AGN is parameterized according to the relative contribution to the total ionizing photon budget
\begin{equation}
f_{\rm AGN} = \frac{Q_{\rm ion, \, AGN}}{Q_{\rm ion, \, AGN} + Q_{\rm ion, \, SSP}} \, ,
\end{equation}where $Q_{\rm ion, \, AGN}$ is the hydrogen-ionizing photon calculated from the AGN spectrum, and $Q_{\rm ion, \, SSP}$ is that from the SSP spectrum.
The input SED to \cloudy\ is simply a weighted sum, constructed according to $f_{\rm AGN}$, and then scaled by $\log(U)$.

\section{Results\label{sec:res}}

In this section, we begin by establishing the connection between the intrinsic spectrum and emission-line strengths using our toy model (\S\,\ref{sec:res:toy}).
We then summarize the key observed features that will be used to constrain the shape of the intrinsic spectrum in LRDs---namely, strong \ha, weak \heii, low \heii/\hb, and a red rest optical continuum (\S\,\ref{sec:res:haheii}--\ref{sec:res:bpt}).
Finally, we present predictions from our \cloudy\ photoionization modeling that account for the presence of dense gas, which introduces complexities in interpreting the emission lines due to reprocessing within the dense medium (\S\,\ref{sec:res:cloudy}).

\subsection{Recombination Lines Probe the Intrinsic Ionizing Spectra\label{sec:res:toy}}

The number of hydrogen-ionizing photons normalized by the continuum flux near \ha\ exhibit a linear relationship with EWs of \ha\ in our toy model (Figure~\ref{fig:toy}).
We emphasize that here we only consider the assumption of ionized gas in an ionization-bounded nebula, and under typical interstellar medium conditions.
Our model clearly demonstrates that the recombination lines probe the integral of the ionizing spectra over a specific wavelength range determined by the ionization potential of the recombination line used, consistent with calculations based on on foundation recombination theory \cite{Ferland2020}.

A similar result holds for \heii. 
However, strictly speaking, the luminosity of the \heii\ line does not always scale linearly with the number of \heii-ionizing photons. 
This nonlinearity stems from microphysical processes involving the relative abundances and ionization states of hydrogen and helium; specifically, the relative availability of \hi-, \hei-, and \heii-ionizing photons. 
\citet{Glatzle2019} show that \hi\ and \hei\ can absorb \heii-ionizing photons at high rates, with \hei\ being especially effective at doing so---more so than \heii\ itself. 
While the exact efficiency depends on factors such as gas composition, their results illustrate the interplay of competing processes that influence the production and escape of \heii\ photons.
The \heii\ response to their ionizing photon budget is still strong, as shown in the bottom right panel of Figure~\ref{fig:toy}.

To further assess the validity of recombination lines as tracers of the intrinsic ionizing spectrum, we compare the predicted \ha\ EWs from our toy AGN model (\S\,\ref{subsec:toy_agn}) to those observed in the SDSS quasar catalog \citep{Paris2018}. The ionizing spectral slope of AGNs is typically found to lie within a relatively narrow range of $\sim -2$ to $-1.2$ in $f_\nu$ vs $\nu$ (or $\sim -0.8$ to $0$ in $f_\lambda$ vs $\lambda$) (e.g., \citealt{Groves2006,Feltre2016}).
Within this range, our toy model predicts \ha\ EWs that are slightly higher than the $+1\sigma$ value of the SDSS quasars.
This is likely due to contamination in the observed EWs due to host galaxy contribution to the continuum, as noted earlier in Section~\ref{sec:data:sdss}. In such cases, the continuum level is overestimated, while the broad line emission flux is not.
Nonetheless, the agreement is broadly consistent and supports the use of recombination lines as tracers for the ionizing properties of the emitting source.

\subsection{Strong H$\alpha$, Weak He\,II, and Red Rest Optical Continua in LRDs\label{sec:res:haheii}}

LRDs as a population tend to exhibit larger \ha\ EWs compared to SDSS AGNs at lower redshifts. This is clearly seen in \rd. 
Under standard conditions, recombination lines such as \ha\ can effectively be treated as ionizing photon counters, implying that whatever powers LRDs is able to produce an abundance of ionizing photons. Moreover, as is illustrated by our toy model (Figure~\ref{fig:toy}), if assuming the power-law ionizing spectra usually used for AGNs, such a high \ha\ equivalent width (EW) implies an ionizing spectrum that is bluer than that of typical AGNs.

At the same time, the observed \heii\ EW is small, which is surprising given the large number of ionizing photons implied by the high \ha\ EW. Even when fit with a broad Gaussian component, the \heii\ EW is only consistent with the values predicted by the softest ionizing spectra expected from AGNs. 
More generally in a larger LRD sample, while the lack of deep medium-resolution spectra precludes sensitivity down to low ($\lesssim 1$~\AA) EWs, the \heii\ emission all appears weak or undetected. 
The stacked spectra exhibit similar features.

This weak \heii\ emission is not unique to \rd. Other deep spectroscopic observations also report weak or absent \heii\ emission in LRDs. Notably, no \heii\ is detected in either individual \citep{Deugenio2025:absorption} or stacked spectra in the JADES survey \citep{Juodzbalis2025}.
A weak \heii\ emission is detected in a local LRD analog at $z=0.1$ \citep{Ji2025:local}, though its flux ratio of \heii/\hb\ is consistent with star-forming galaxies in the local Universe.

Adding more to the puzzle, though the strong \ha\ emission observed in LRDs imply an intrinsically blue spectrum, the observed rest optical continua of LRDs are much redder than those of typical AGNs.
This discrepancy is illustrated in Figure~\ref{fig:prism}, where the LRD spectra deviate significantly from standard AGN templates.

We discuss possible interpretations of these apparent contradictions in the observed features in Sections~\ref{sec:dis:turnover} and \ref{sec:dis:reprocess}.

\subsection{He\,II/H$\beta$ Diagnostic\label{sec:res:bpt}}

\begin{figure} 
\gridline{
  \fig{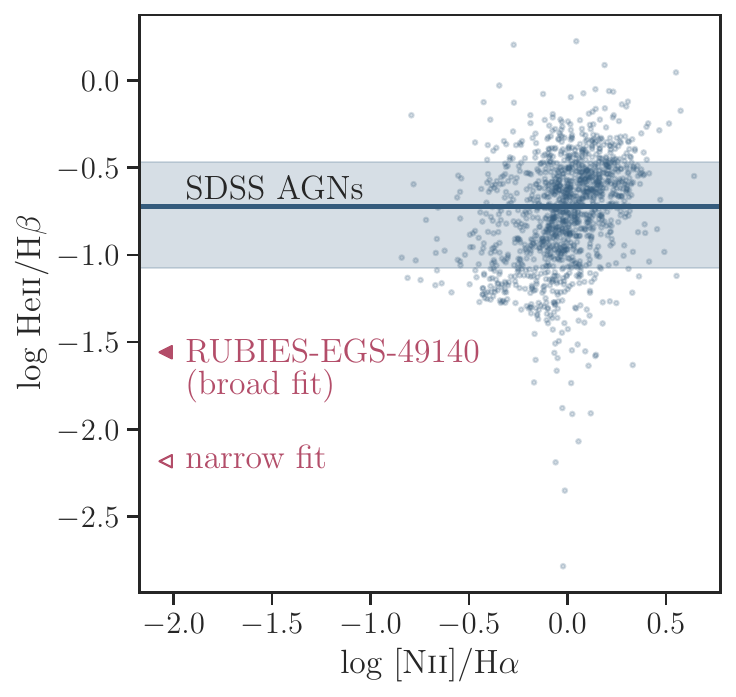}{0.45\textwidth}{}
}
\caption{\heii-based diagnostic for AGNs. SDSS AGNs are shown as blue points. The median \heii/\hb\ ratio of $10^{-0.7}$ is indicated by a horizontal blue line, with the $\sim0.3$ dex $1\sigma$ scatter shaded in blue.
For \rd, this \heii/\hb\ line ratio measured using a narrow Gaussian model for \heii\ is shown as a red unfilled triangle, while the measurement including an additional broad \heii\ component is shown as a red triangle. \nii\ is undetected in the medium-resolution G395M spectrum, and so is reported as an upper limit using the 95th percentile. In both cases, the \heii/\hb\ values lie well below the typical range of AGNs at $z\sim0$. Combined with the low \nii/\ha\ ratio, \rd\ falls securely within the star-forming region of the \heii-based diagnostic diagram.}
\label{fig:heiihb}
\end{figure}

We confirm the intuition regarding the surprisingly soft ionizing spectrum from the previous section from \heii\ EWs by examining the \heii/\hb\ line ratio diagram. This diagnostic is largely invariant to the gas conditions, and thus more sensitive to the shape of the ionization spectrum, compared to the conventional BPT \citep{Baldwin1981} emission-line diagnostic \citep{Shirazi2012}.
In a classic BPT diagram (\oiii/\hb\ versus \nii/\ha), metal-poor stars can produce line ratios that resemble those of AGNs, making it difficult to distinguish between the two ionizing sources (see \citealt{Kewley2019} for a review, and e.g., \citealt{Cameron2023,Sanders2023,Cleri2025}). However, the ionization potential of He$^{++}$ (54.4 eV) is significantly higher than that of O$^{++}$ (35.2 eV) and N$^+$ (15.5 eV), making this diagnostic less susceptible to contamination.
In Figure~\ref{fig:heiihb}, we show the SDSS AGNs on the plane of \heii/\hb\ vs. \nii/\ha, as compiled by \citet{Bar2017}.

The velocity resolution of the medium-resolution grating is, in principle, sufficient to resolve \nii. However, in \rd, the \ha\ line is so strong that \nii\ cannot be meaningfully constrained. We therefore adopt the 95th percentile as an upper limit, obtaining a low \nii/\ha\ ratio of $<10^{-2}$, consistent with previous findings for JWST AGN samples (e.g., \citealt{Juodzbalis2025}).

Most SDSS AGNs exhibit high \heii/\hb\ ratios, as naturally expected from hard ionizing spectra.
The median \heii/\hb\ of the \citet{Bar2017} sample is $10^{-0.7}$, with a $1\sigma$ scatter of about $0.3$~dex.
In contrast, we measure extremely low \heii/\hb\ values for \rd: $10^{-2.2}$ when modeling \heii\ with a narrow Gaussian component, and $10^{-1.6}$ when including an additional broad component. In both cases, the values fall well below the typical range observed in AGNs at $z \sim 0$.
This discrepancy may, in fact, be even more severe. While multiple studies suggest that LRD spectra contain little dust, their Balmer decrements deviate from the values expected under Case B recombination, indicating that collisional de-excitation of \hb\ may be occurring; and thereby reducing the emissivity (e.g., \citealt{Yan2025}).
Combined with the low \nii/\ha\ ratio, \rd\ falls securely within the star-forming region of the \heii-based diagnostic diagram.

In summary, the line-ratio analysis, particularly the exceptionally low \heii/\hb\ reported here, supports the conclusion from the EW analysis: the implied ionizing spectrum of LRDs is much softer than typical AGNs.

\subsection{Revisiting the Emission-line Diagnostics in the Presence of Dense Gas\label{sec:res:cloudy}}

The inference of intrinsic ionizing spectra of LRDs becomes less certain with the black hole star hypothesis \citep{Inayoshi2025:dense,Ji2025,Naidu2025,deGraaff2025:cliff}. If the central ionizing source is embedded in a cloud of dense hydrogen gas, the emitted radiation may be so thoroughly reprocessed that the emergent lines no longer directly connect to the intrinsic spectrum.
In what follows, we assess if and how such extreme physical conditions alter our proposed recombination line diagnostics.

\subsubsection{He\,II/H$\beta$ Diagnostic}

\begin{figure*}
\gridline{
  \fig{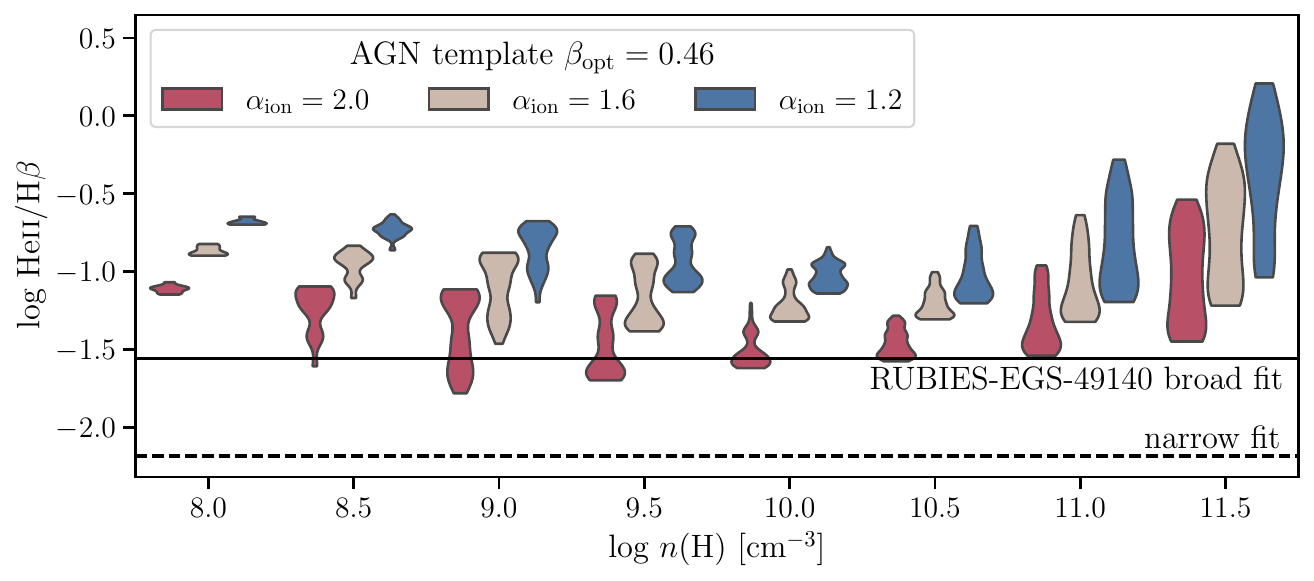}{0.9\textwidth}{(a)}
}
\gridline{
  \fig{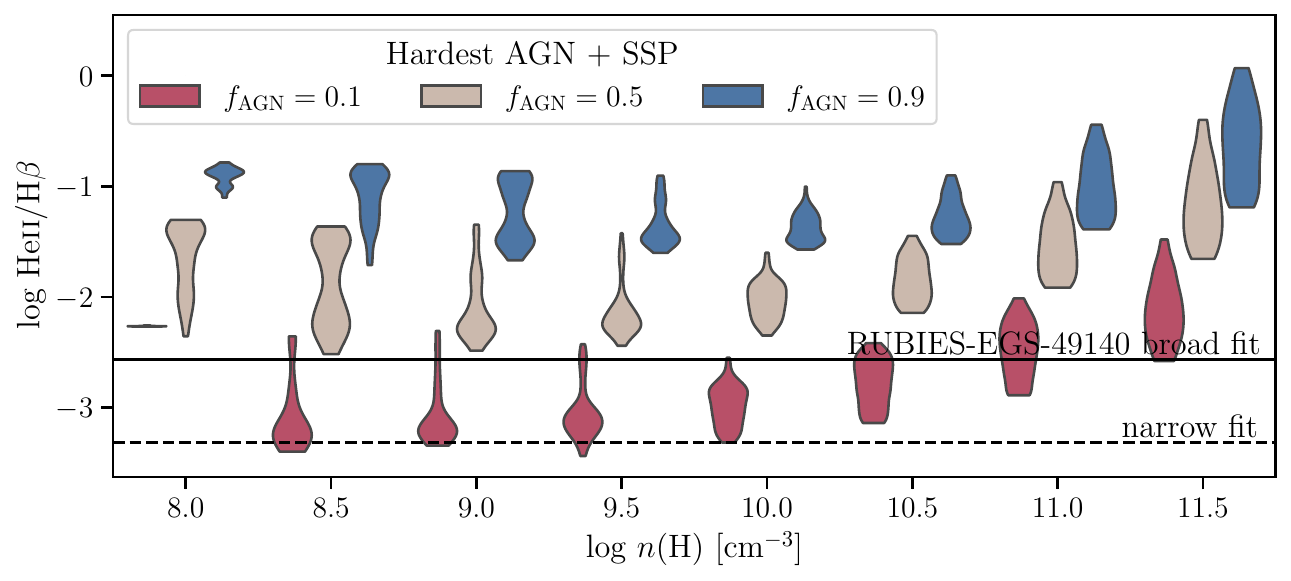}{0.9\textwidth}{(b)}
}
\caption{\heii /\hb\ flux ratio as a function of number density of neutral hydrogen for two sets of \cloudy\ models: (a) fiducial AGN template with varying optical continuum slope, $\bopt$; (b) AGN-stellar composite models with various levels of AGN contributions to the ionization photon budget, $\fagn$. The patches show violin plots with symmetrical axes, representing the distributions of the line ratios marginalized over all free parameters beside $n_{\rm{H}}$ in our \cloudy\ models.
The exceptionally low values of \heii /\hb\ of \rd, plotted as horizontal lines, suggest that its intrinsic SED does not resemble that of a typical AGN.}
\label{fig:heiihb_nh}
\end{figure*}

We start by examining whether the \heii/\hb\ line ratio diagnostic, as studied in Section~\ref{sec:res:bpt}, remains effective when the ionizing source is embedded in dense neutral hydrogen. Figure~\ref{fig:heiihb_nh} shows the \heii/\hb\ ratio as a function of neutral hydrogen number density, $n_{\rm{H}}$, based on our \cloudy\ model grid for two cases.

First, we show the predicted line ratios assuming the fiducial AGN template with varying ionizing spectral slopes, from the hardest $\aion = 1.2$ to the softest $\aion=2$ typical of AGNs.
Notably, at low densities ($n_{\rm{H}} \lesssim 10^{8} \, {\rm cm^{-3}}$), the \heii/\hb\ flux ratio appears to retain sensitivity to the hardness of the ionizing spectra, with clear separation between the model predictions. As $n_{\rm{H}}$ increases, however, its diagnostic power diminishes for reasons examined in the next subsection.
If assuming a narrow \heii\ emission, the measured values of \rd\ remain significantly below those predicted for AGNs across the full range of densities, whereas if assuming \heii\ have a broad component, the measured flux ratio is only consistent with values expected from standard AGNs with the softest ionizing spectra at densities ($n_{\rm{H}} \gtrsim 10^{8.5} \, {\rm cm^{-3}}$).

Second, we consider composite spectra including both AGN and stellar contributions as incident SEDs. 
This can be thought of as simply a way to soften the ionizing SED. The potential physicality of this will be discussed later in Section~\ref{sec:comp}.
Interestingly, the observed \heii/\hb\ ratios in \rd\ are consistent with models that include substantial stellar contributions ($\fagn \lesssim 0.5$). Models with high AGN fractions ($\fagn \sim 0.9$) consistently yield \heii/\hb\ $\sim 10^{-1}$, whereas models with low AGN fractions ($\fagn \sim 0.1$) predict much lower ratios ($\sim 10^{-3}$).
However, as the density increases to values usually required to fit LRD spectra ($n_{\rm H} \sim 10^{10} - 10^{11}~{\rm{cm^{-3}}}$; \citealt{deGraaff2025:cliff,Ji2025,Naidu2025}), these distributions begin to shift and overlap. This trend critically limits the utility of the \heii/\hb\ ratio as a robust diagnostic.

\subsubsection{Origin of Line Emission within Dense Gas\label{sec:rad}}

\begin{figure*} 
\gridline{
  \fig{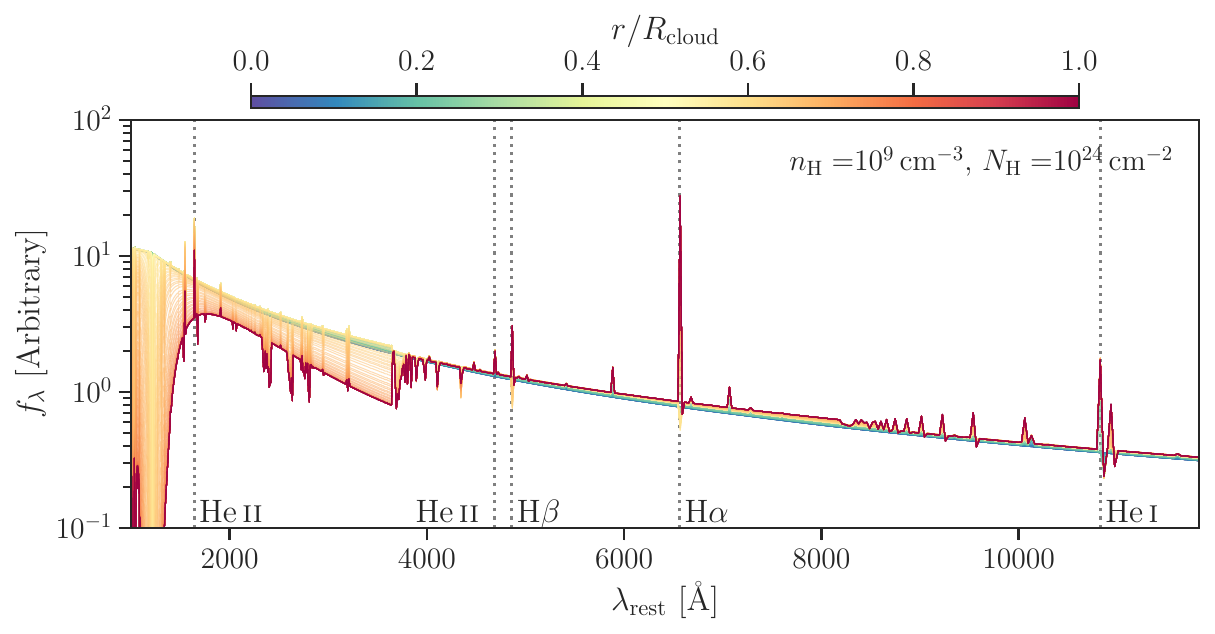}{0.9\textwidth}{(a)}
}
 \gridline{
   \fig{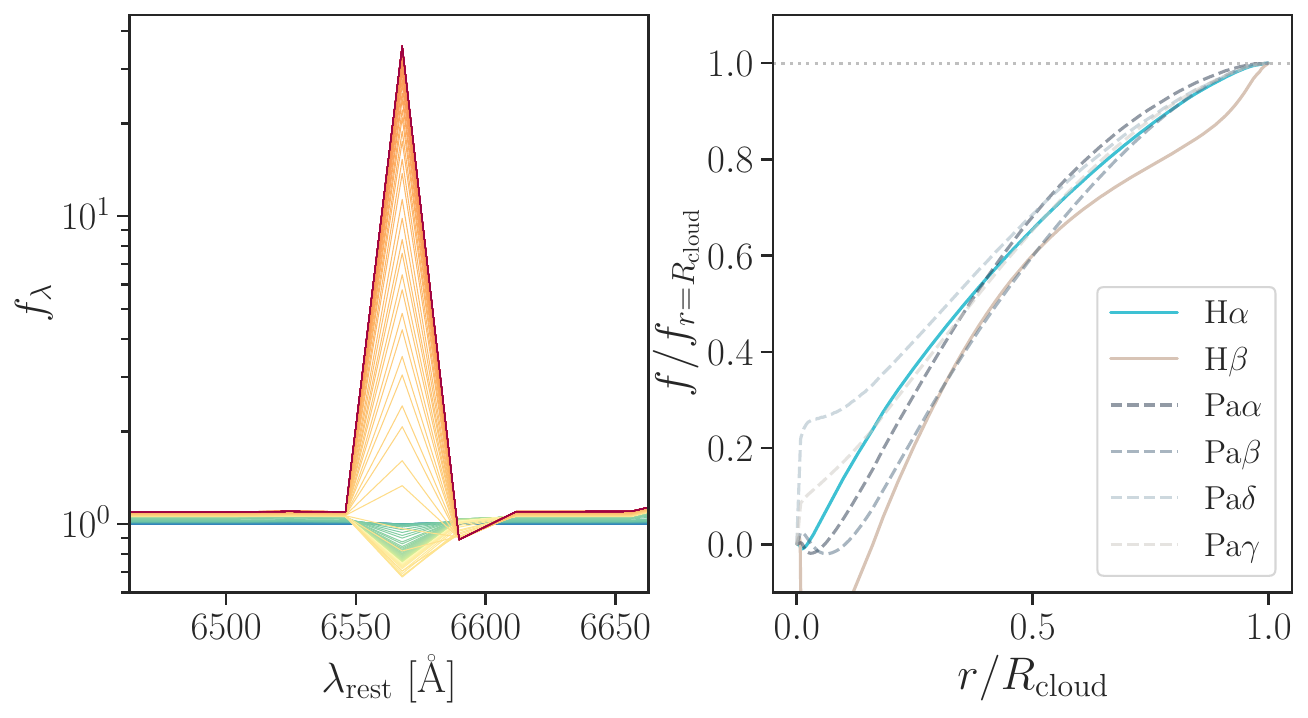}{0.45\textwidth}{(b)}
   \fig{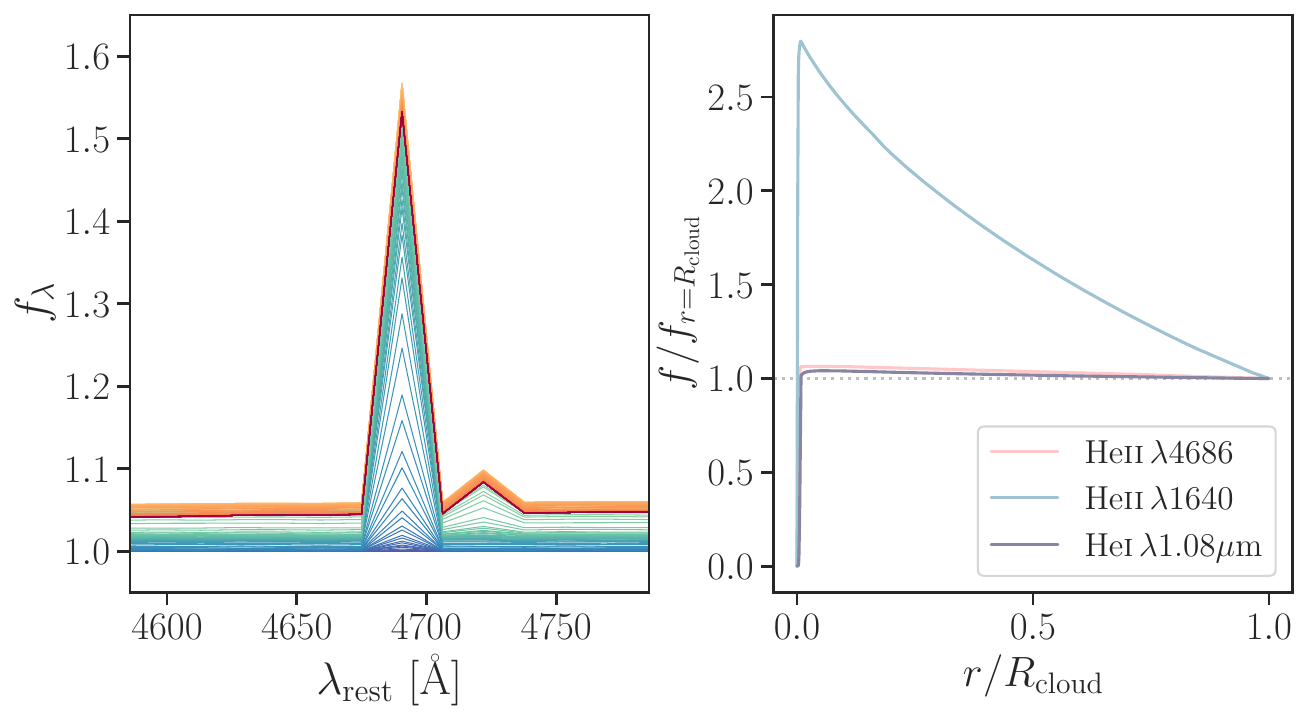}{0.45\textwidth}{(c)}
}
\caption{Origins of emission line fluxes. (a) The net transmitted spectra are shown at various cloud radii, from the core in blue ($r/r_{\rm cloud}=0$) to the outer edge in red ($r/r_{\rm cloud}=1$). (b) Hydrogen line fluxes normalized by the emergent line flux as a function of cloud radii. (c) Helium line fluxes normalized by the emergent line flux as a function of cloud radii. In all panels, the values shown at each radius correspond to line-of-sight integrations of the respective quantities of SED and line fluxes.
The hydrogen recombination lines are primarily produced near the outer edge of the cloud, whereas the helium recombination lines are predominantly generated near the core of the cloud.}
\label{fig:rad}
\end{figure*}

\begin{figure*}
\gridline{
  \fig{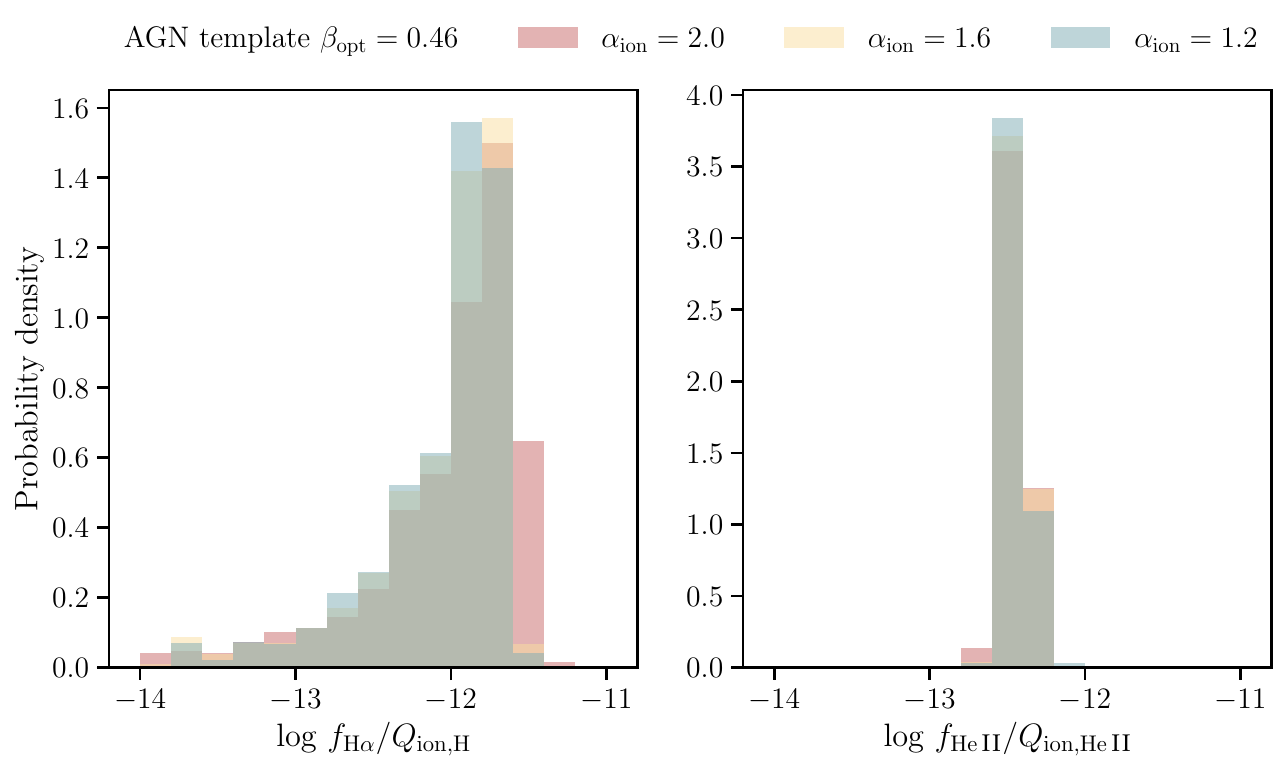}{0.9\textwidth}{}
}
\caption{Recombination line flues per ionizing photon, from incident SEDs assuming the fiducial AGN template with varying ionizing spectral slops, $\aion$. (a) The distributions of \ha\ flux per hydrogen-ionizing photon $\qion$ exhibit wide spreads, suggesting that \ha\ gets processed in the dense gas, and thus the expected physical relationship between the line flux and ionizing photons vanishes. (b) \heii\ flux per helium-ionizing photon $\qionhe$, in contrast, shows much narrower spreads, supporting our finding that \heii\ retain information about the intrinsic ionizing spectra, even after passing through a cloud of dense gas.}
\label{fig:cloudy_ew}
\end{figure*}

To understand the changes in the \heii/\hb\ line ratios due to the addition of dense neutral hydrogen, we must examine the origin of line emission within the gas.
The net transmitted spectra as the incident radiation propagates through a dense hydrogen cloud are shown in Figure~\ref{fig:rad}. In the same figure, we highlight the formation of the \ha\ and \heii\ lines as a function of radius, and also show the line fluxes normalized by the emergent flux (i.e., the flux seen outside the cloud) for a representative set of lines.

It is evident that the hydrogen recombination lines are primarily produced near the outer edge of the cloud. This suggests that the radiation is heavily reprocessed by the intervening gas and emerge only after it has traveled through the bulk of the cloud.
Moreover, the Balmer lines are first appeared in absorption, and then in emission as the radiation passes through the cloud, further supporting the finding that the gas cloud is optically thick to Balmer line emission.
Consequently, the emergent hydrogen recombination line fluxes are no longer directly tied to the central ionizing source.

In contrast, helium recombination lines are predominantly generated near the core of the cloud. This means that the cloud is optically thin to \heii\ emission and the connection to the central engine is maintained.
An addition implication is that the helium lines are expected to be broader since they are generated closer in.
It is worth pointing out, however, that the strong gas attenuation blue-ward of the Balmer break means that the UV lines such as \heiiuv\ are also heavily attenuated in the models and commonly appear in absorption, and thus not suitable for diagnostics.

Taking a step further, we examine the line fluxes normalized by the number of corresponding ionizing photon.
Figure~\ref{fig:cloudy_ew} shows that the distributions of \ha\ flux per hydrogen-ionizing photon, $\qion$, have wide spreads spanning $\sim 2$~dex.
However, the \heii\ flux per helium-ionizing photon, $\qionhe$, exhibits much narrower spreads, with minimal variation across models with different incident SEDs. 
Explanations and implications from these findings are discussed in Section~\ref{sec:dis:reprocess}.

\subsubsection{He\,II EW\label{sec:res:ew}}

\begin{figure*} 
\gridline{
  \fig{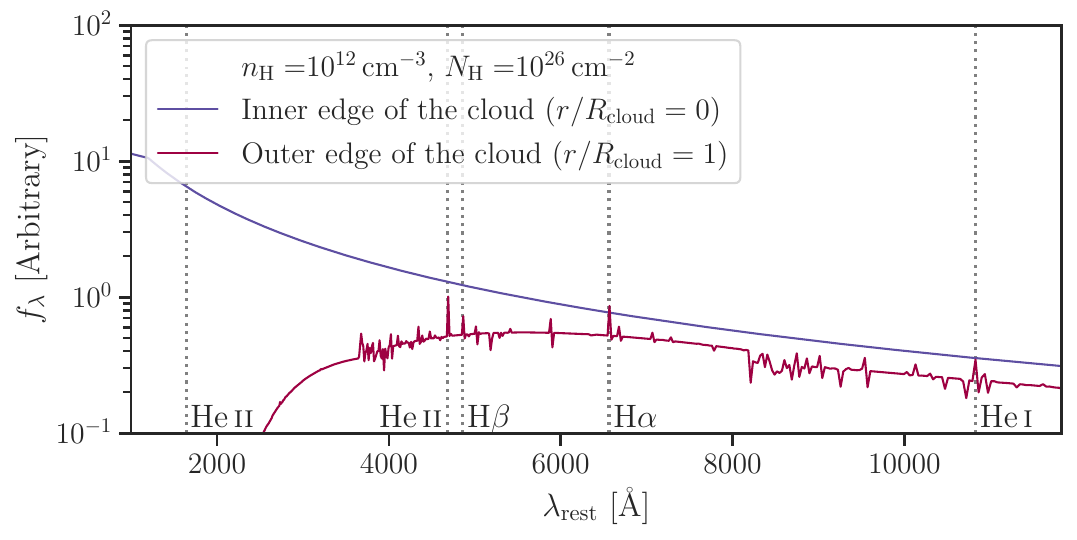}{0.9\textwidth}{(a)}
  }
\gridline{
  \fig{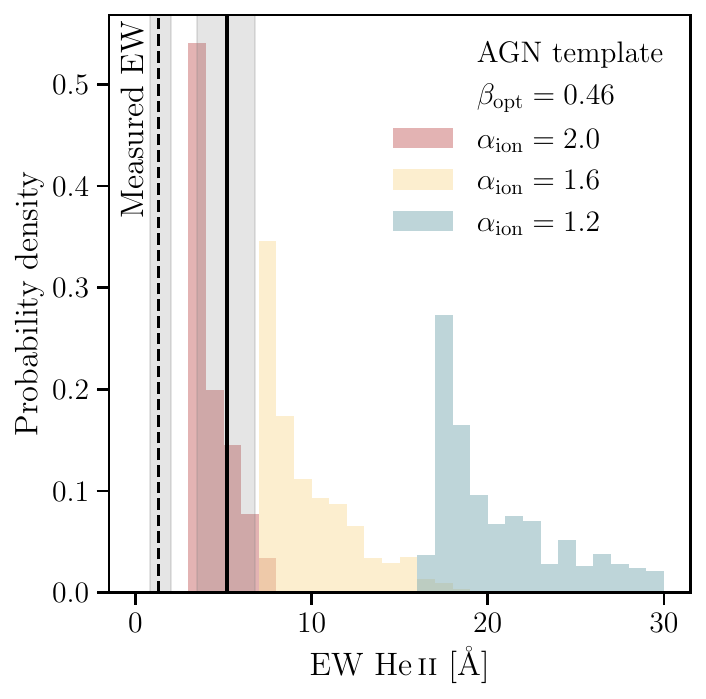}{0.45\textwidth}{(b)}
  \fig{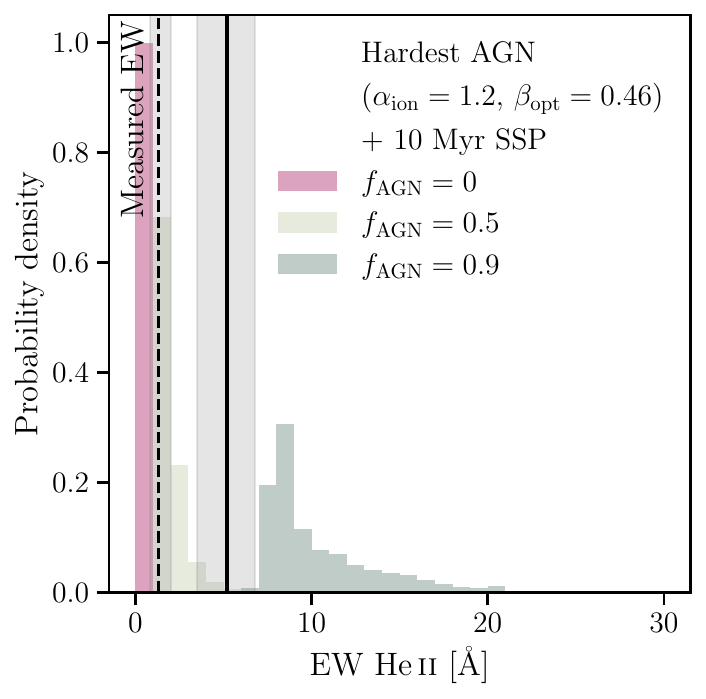}{0.45\textwidth}{(c)}
 }
\caption{Observers measure EWs, which introduces a potential dependence on the shape of the continuum. (a) Similar to Figure~\ref{fig:rad}, but the neutral hydrogen density is increased here to show the absorption of the continuum.
(b) Histograms show the \heii\ EWs from the model grid consisted of a fiducial AGN incident SED with varying ionizing spectral slopes, $\aion$. Measured \heii\ EWs of \rd\ assuming the narrow and the broad models are plotted as a vertical dashed line and solid line, respectively. The shading corresponds to the 1$\sigma$ uncertainty.
(c) The same, but for the model grid consisted of composite AGN and stellar spectra with varying AGN contributions to the total ionization photon budget, $\fagn$.}
\label{fig:rad_dense}
\end{figure*}

In observations, the most straightforward normalization-independent metric is emission line EW, which introduces a potential dependence on the shape of the continuum near the \heii\ line. 
At modest optical depths at this wavelength, the intrinsic continuum is absorbed but may not be completely reprocessed into blackbody emission; at higher optical depths, however, the continuum is simply a blackbody.
We note that blackbody curves are found to describe LRDs well (e.g., \citealt{Inayoshi2025:bbh}), suggesting that the latter is often the case.

Despite this added complexity, \heii\ EW remains useful in distinguishing different ionizing sources. In Figure~\ref{fig:rad_dense}, we show histograms of the \heii\ EW distributions from a set of representative intrinsic SEDs.
Our fiducial AGN model with the hardest ionizing spectrum predicts \heii\ EWs in the range of 14 -- 44~\AA, with a median of 20~\AA, $-1\sigma=2$~\AA, and $+1\sigma=8$~\AA. In contrast, the \heii\ EWs predicted by the same model but with the softest ionizing spectrum are much lower, spanning the range of 3 -- 9~\AA. The median ($-1\sigma$) ($+1\sigma$) is 3.9~\AA\ ($-0.5$) $(+1.8$).

In Section~\ref{sec:line}, the \heii\ EW of \rd\ is measured to be $1.3\pm0.6$~\AA\ under the assumption of a narrow Gaussian line profile. This measurement thus confidently rules out a standard AGN with the hardest expected ionizing spectrum ($\aion=1.2$) as the central ionizing source of LRD. Even the softest AGN ionizing spectrum ($\aion=2.0$) is excluded at a $3\sigma$ level.

Conservatively, we also measure the \heii\ EW of \rd\ assuming a broad Gaussian model, obtaining an EW of $5.2^{+1.6}_{-1.7}$~\AA. This measurement still rules out a standard AGN with $\aion = 1.2$, but is consistent within $1\sigma$ with a standard AGN having the softest ionizing spectrum.

The low \heii\ EWs observed in LRDs motivate us to explore additional models with soft ionizing spectra. In particular, composite spectra combining a hard AGN component with a SSP may offer further insight. Interestingly, the narrow \heii\ EW is consistent within $1\sigma$ of the prediction from a 10~Myr, sub-solar metallicity SSP. On the other hand, the broad \heii\ EW is best described with a composite spectrum of $\fagn = 0.5$, and remains consistent within $3\sigma$ with one dominated by a high AGN contribution (e.g., $\fagn = 0.9$).

\subsection{Joint Constraints from HeII EW and Rest Optical Continuum Slope\label{sec:res:joint}}

\begin{figure*}
\gridline{
  \fig{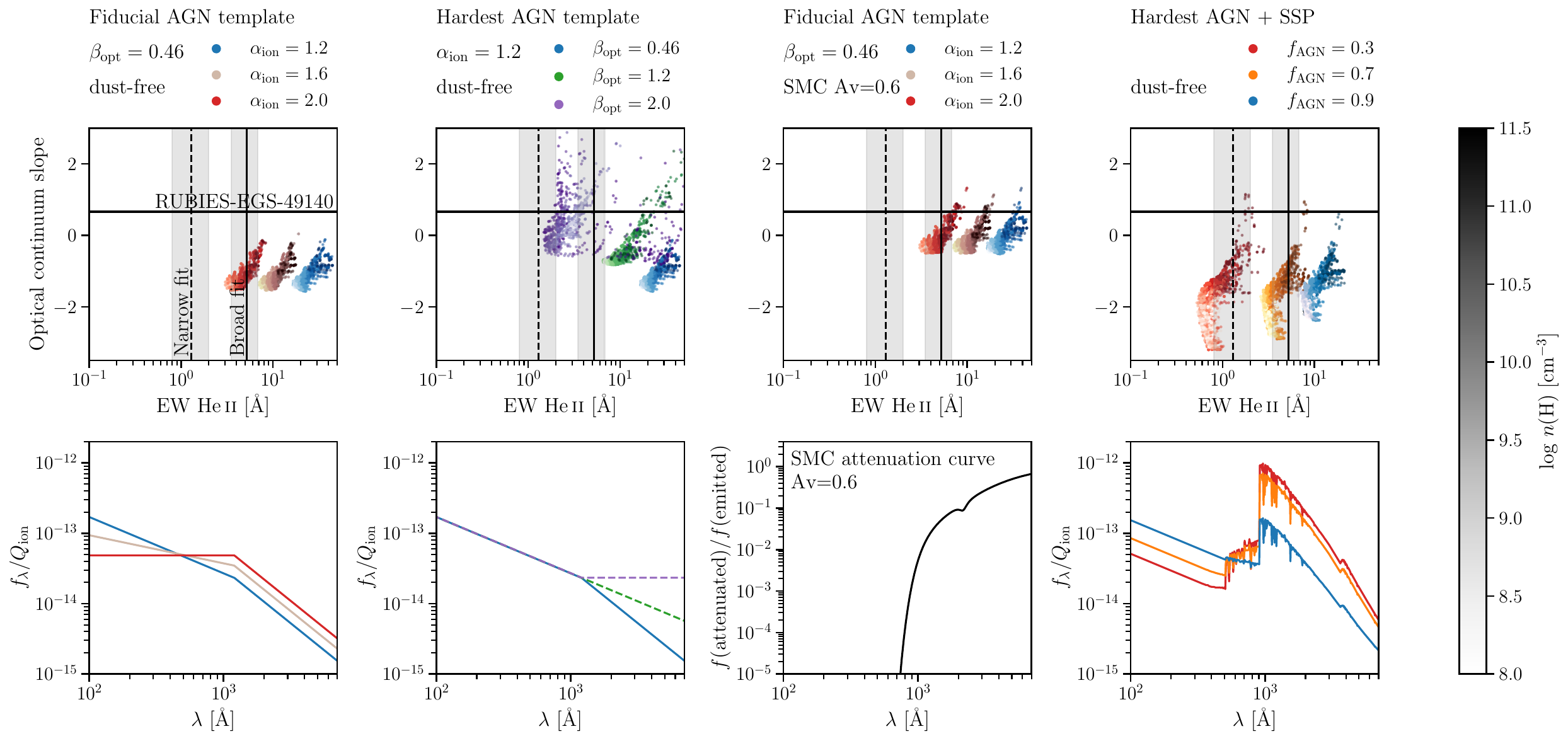}{0.99\textwidth}{}
}
\caption{Various physical scenarios and comparisons to the observed features of \rd. From left to right, we show the results from four model sets: dust-free fiducial AGN template with the ionization spectral slope varying within the expected range of typical AGNs, dust-free AGN template with the hardest ionizing spectrum with varying optical continuum slopes, same as the fiducial AGN model but attenuated with a small amount of dust, and composite AGN and stellar spectra with varying AGN contributions to the total ionization photon budget. The first row shows the model predications on the plane of \heii\ EW and optical continuum slope. The color scale indicates number density of neutral hydrogen, and the measured values of \rd\ are plotted as black lines. The second row illustrates the incident SEDs in the dust-free models, and the SMC attenuation curve used in the dusty models.
}
\label{fig:slope}
\end{figure*}

Having thus far focused primarily on constraining the ionizing spectrum, we now turn to possible constraints on the full SED shape of LRDs.
The presence of a dense gas envelope links the optical slope, the observed \ha\ EWs, and the strength of the spectral break, all of which are determined by properties of the hydrogen envelope including the density, temperature, and column depth.
The results are summarized in Figure~\ref{fig:slope}, which shows the optical continuum slope that we use as a proxy for redness (\S\,\ref{subsec:bopt}) as a function of \heii\ EW for four sets of \cloudy\ models. 

We dedicate this section to present results from each model. 
While our comparisons to model predications are made specifically with \rd, given that it is one of the brightest LRDs with high-quality deep spectroscopy enabling a more precise measurement of the \heii\ EW, our conclusions are expected to hold more broadly across the LRD population.
This is because \heii\ appears to be similarly weak in both Prism and shallow grating spectra for a larger sample of LRDs. Moreover, \rd\ belongs to the RUBIES LRD sample, selected uniformly based on a red optical continuum ($\bopt > 0$), unresolved morphology in the optical, and broad Balmer lines \citep{Hviding2025}. These shared spectral characteristics suggest that the physical interpretations drawn from \rd\ are likely representative of the class as a whole.

\subsubsection{Fiducial AGN Model without Dust: Inconsistent with LRDs\label{subsec:res:fid}}
 
At high neutral hydrogen densities ($n_{\rm{H}} \gtrsim 10^{11} \, {\rm cm^{-3}}$), the dense gas reprocesses the continuum light, so that an intrinsically blue optical continuum becomes redder, approaching a slope of $\bopt \sim 0$. However, typical LRDs exhibit even redder continua.

The first column of Figure~\ref{fig:slope} clearly shows that standard AGN SEDs fail to reproduce the region of parameter space occupied by \rd; namely, a red optical color and low \heii\ EW. The same holds when adopting the AGN templates from \cite{Jin2012}. Among them, the SED corresponding to the lowest Eddington ratio has the softest ionizing spectrum and predicts \heii\ EWs of $6-17$~\AA. It is the only template capable of producing a low \heii\ EW that is consistent with the observed values. However, its rest-optical continuum is similarly too blue, and thus inconsistent with the observed LRD continua.

There are three variables that could affect the model predictions and potentially satisfy the two observational constraints. We explore each of these possibilities in the following subsections.

\subsubsection{Varying the Rest-Optical Continuum Slope\label{subsec:res:opt}}

An obvious alternative is to make the slope of the rest-optical continuum of the incident SED redder. The second column of Figure~\ref{fig:slope} shows the changes in model predictions when fixing the ionizing spectral slope and varying the optical slope $\bopt$. To illustrate this effect, we use the hardest ionizing spectrum in our model grid. The redder optical continua (i.e., higher $\bopt$) also lead to lower \heii\ EWs, primarily due to the elevated continuum level.

This result suggests that, with an intrinsically redder rest-optical SED---such as one that is nearly flat in $f_\lambda$---even the hardest ionizing spectra typical of AGNs can simultaneously reproduce both the low \heii\ EW and the red optical color observed in LRDs, without invoking dust attenuation. 
However, as alluded to in \ref{subsec:bopt}, we emphasize again that there is no obvious physical mechanism that would naturally produce such a red intrinsic spectrum assuming a standard AGN, where the emission from the accretion disk is a single-temperature blackbody peaking at UV wavelengths. A more physically motivated analysis would benefit from adopting the super-Eddington accretion models recently proposed by \citet{Liu2025}.

\subsubsection{AGN Model with Modest Dust Attenuation\label{subsec:res:dust}}

Since high neutral hydrogen densities ``redden'' the continuum, introducing a modest amount of dust attenuation can further increase $\bopt$, reconciling the predicted \heii\ EW and optical color with those observed in LRDs.
While we do not attempt to fully explore the broad parameter space opened by including dust, we provide an illustrative example using the best-fit dust model from \citet{deGraaff2025:cliff,Naidu2025}: an SMC attenuation curve with $\Av = 0.6$.

As shown in the third column of Figure~\ref{fig:slope}, applying this attenuation shifts all predicted $\bopt$ upward into the regime occupied by LRDs. However, more dust, or a steeper dust attenuation curve may be needed to bring the model predications into full consistency with the measurements of \rd.
To do this with dust alone requires $\Av \gtrsim 2$, which is currently ruled out on average for the sample by \citet{Akins2025:alma,Casey2025}, and by deep observations of a few individual objects \citep{Setton2025:alma,Xiao2025}. A steep dust attenuation curve is likewise unlikely \citep{deGraaff2025:cliff,Ma2025}. The dusty AGN interpretation is also directly challenged by the recent discovery of molecular absorption features in $z \sim 2$ LRDs: simple temperature arguments show that such molecules would be completely dissociated at the temperatures characteristic of typical AGN accretion disks \citep{Wang2026:h2o}.

While applying dust attenuation as a post-processing step is common practice in SED modeling, we note that including dust directly in \cloudy\ calculations introduces additional physical effects due to photoelectric heating from dust grains. For ionizing conditions typical of star-formation-dominated regions (e.g., H~\textsc{ii} regions), the inclusion of dust has only a minor impact on the \heiiir\ emission, at the level of $\sim 10$\%. In AGN-like conditions, dust can in principle contribute more significantly through enhanced photoelectric heating, as predicted by photoionization models. A full exploration of dust physics is beyond the scope of this paper. However, based on a small set of tests, we find that this effect does not qualitatively alter our results: crucially, the key insight of this work---that the \heii\ emission originates near the core of the gas cloud and retains information about the intrinsic ionizing spectra---remains valid.

\subsubsection{Composites of AGN and Stars without Dust\label{subsec:res:stars}}

Adding a stellar component to the AGN SED alters both the ionizing spectrum and the rest-optical continuum shape. As noted in Section~\ref{sec:res:ew}, the inclusion of stellar light softens the ionizing spectrum, leading to lower \heii\ EW. In Figure~\ref{fig:slope}, we further show that models with low to modest AGN fractional contributions ($\fagn<0.7$) can reproduce both the red optical colors and the low \heii\ EW observed in LRDs, in the absence of dust attenuation.

Tying the stellar and gas-phase metallicities yields qualitatively similar results. The only noticeable effect is increased scatter in the plane of \heii\ EW versus optical continuum slope. The increased scatter is expected since this choice effectively introduces additional variations in the input spectra.

\section{Discussion\label{sec:dis}}

In this section, we begin by discussing the standard case without the presence of dense gas, where the interpretation of recombination lines is more straightforward. 
This remains a useful exercise, given the lack of a definitive physical model for LRDs and the possibility that they constitute a heterogeneous population. These findings may apply to some of the LRDs with lower gas column densities.
In such conditions, the rest-optical continuum of AGNs normally exhibits minimal variation. We thus focus our discussion on constraints on the ionizing spectrum of LRDs, and highlight the peculiar shape of their ionizing spectra suggested from these constraints (\S\,\ref{sec:dis:turnover}--\ref{sec:dis:heiihb}).

We then proceed to examine how the presence of dense gas alters our interpretations. In high-density environments, both line and continuum emission can be reprocessed, introducing complications for inferring the intrinsic SEDs of LRDs.
Such reprocessing includes absorption and re-emission, scattering, and collisional excitation and de-excitation (i.e., thermalization).
Notably, the reprocessing goes beyond simple absorption and re-emission, which would otherwise conserve flux. The deviation from the Balmer decrement values expected under Case B recombination without dust in the models strongly suggests the occurrence of thermalization, consistent with hints from observed Balmer decrements in LRDs. 
The effects from the reprocessing on interpreting the emission lines, and the implied intrinsic SEDs, are discussed in the second half of this section (\S\,\ref{sec:dis:reprocess}--\ref{sec:dis:joint}).

\subsection{An Ionizing Spectrum with a Peculiar Turnover at the Helium-ionizing Wavelength\label{sec:dis:turnover}}

LRDs tend to exhibit large \ha\ EWs, often exceeding $500$~\AA\ (e.g., \citealt{Matthee2024,Lin2024,Hviding2025}), and overall show systematically higher EWs than those observed in standard AGNs (e.g., \citealt{VandenBerk2001,Greene2005,Stern2012}).
As clearly demonstrated by our toy model assuming only photoionization, such large \ha\ EWs mean large values of $Q_{ion}$ over the continuum flux at \ha.
Under the assumption of a single-slope power-law spectrum, large \ha\ EWs imply an intrinsically very blue ionizing spectrum, corresponding to a UV slope with $\aion < -1$.
Given that the AGN ionizing spectra are conventionally taken to follow a power-law form since it is generally the tail of a blackbody, this leads to the first major puzzle: such a hard ionizing spectrum would typically also produce strong \heii\ emission, resulting in a large \heii\ EW. However, this is not observed.

Taken at face value, this discrepancy suggests that the ionizing spectrum in LRDs has a peculiar shape---one that produces an abundance of hydrogen-ionizing photons (energies $>$ 13.6~eV; $\lambda<$ 912~\AA), but relatively few helium-ionizing photons (energies $>$ 54.4~eV; $\lambda<$ 228~\AA). In other words, the ionizing continuum must be blue enough to explain the observed \ha\ emission, yet soft enough to suppress \heii\ emission, pointing to an atypical or composite ionizing source.

Adding more to the puzzle, reconciling an intrinsically blue SED with a red observed slope requires significant dust attenuation.
Indeed, this has been a key assumption in early modeling efforts of LRDs \citep{Wang2024:ub,Wang2024:brd,Labbe2024:monster,Ma2025}, which successfully explain many observed features. However, this interpretation comes into tension with the flat MIRI detections (\citealt{Wang2024:brd,deGraaff2025:cliff}; also see \citealt{Perez-Gonzalez2024,Williams2024,Akins2024}), as well as with the non-detected dust continuum emission in ALMA \citep{Akins2025:alma,Casey2025,Setton2025:alma,Xiao2025}.

These recent findings point toward little or no dust, or the need for alternative dust models such as extended dusty outflows \citep{Li2025}.
Under the standard assumption, explaining both the large \ha\ EWs and the red optical continuum shape becomes difficult if the intrinsic spectrum resembles that of a typical AGN.

\subsection{A Non-AGN Intrinsic Spectrum Implied by HeII-based Line Ratio Diagnostic\label{sec:dis:heiihb}}

In addition to inferring the ionization properties from our toy model, we also examine commonly used line ratio diagnostics for distinguishing AGNs from star-forming galaxies. In particular, we focus on the \heii-based diagnostic proposed by \citet{Shirazi2012}.

Typical AGNs are rarely observed to have log \heii/\hb\ $<-1$ \citep{Shirazi2012,Bar2017}, with composite systems primarily populate the region around log \heii/\hb\ $\sim -1$. As shown in Figure~\ref{fig:heiihb_nh}, the \heii/\hb\ measured for \rd\ is significantly lower than the typical values seen in SDSS AGNs. 

While some SDSS AGNs are outliers having low \heii/\hb\ ($\lesssim 10^{-1.5}$), these sources tend to exhibit elevated \nii/\ha, indicative of super-solar metallicities (e.g., \citealt{Groves2006}). 
Notably, \nii\ is undetected in the 8~hr medium-resolution spectrum of \rd, placing an upper limit on \nii/\ha\ $<10^{-2}$. With such a low \nii/\ha\ flux ratio, \rd\ is securely located in the star-forming region of the line ratio diagram.

Together, these results suggest that the intrinsic ionizing spectrum of LRDs is unlikely to be consistent with that of standard AGN. Instead, it appears to be much softer, resembling more of an ionizing spectrum with significant stellar contributions ($\fagn<0.7$), or a much colder accretion disk (e.g., \citealt{Davis2011,Laor2014}).

We opt to compare to the SDSS AGNs in order to make empirical arguments, although we note that \citet{Nakajima2022} proposed to use \heii\ EW and \heii/\hb\ flux ratio as a diagnostic for distinguishing among direct collapse black holes (DCBHs), Population III, and Population II galaxies in pristine environments. In their model grid, \rd\ falls roughly in the region occupied by the DCBHs. 
These black holes are theorized to form from the collapse of primordial massive gas clouds, under the condition where sufficient Lyman-Werner photons are present to suppress gas fragmentation. DCBHs are expected to have masses in the range of $10^5 - 10^6~\msun$. 
Such systems have long been proposed as one pathway for building up massive black hole seeds at $z \sim 10-15$ (see \citealt{Volonteri2012,Inayoshi2020} for reviews, and e.g., \citealt{Bromm2003,Shang2010,Johnson2012,Ferrara2014,Valiante2016}), and have received renewed attention, as early JWST results indicate galaxies host overly massive black holes (e.g., \citealt{Goulding2023,Furtak2024}). Recent work suggests that DCBH models can reproduce the observed demographics and properties of LRDs \citep{Cenci2025,Jeon2025}.

Interestingly, the \heii/\hb\ line diagnostic remains useful when the ionizing source is embedded behind a neutral hydrogen cloud, provided that the number density is not too high ($n_{\rm H} \lesssim 10^8~{\rm cm^{-3}}$). At higher densities, which are typically required to fit the LRD spectra, the distributions of \heii/\hb\ from various ionizing spectra begin to overlap and show systematic shifts (Figure~\ref{fig:heiihb_nh}).
This behavior is primarily driven by the reprocessing of \hb\ within the dense medium. It is worth pointing out that collisional de-excitation becomes important in this high-density regime: a fraction of excited H$^+$ de-excites without producing Balmer emission, such that ionizing photons absorbed by neutral hydrogen are no longer fully traced by \hb. As a result, the \heii/\hb\ ratio no longer reliably traces the intrinsic ionizing spectrum.
We elaborate on this point in the following subsections.

\subsection{Reprocessing of Recombination Lines in Dense Neutral Hydrogen Gas\label{sec:dis:reprocess}}

In Section~\ref{sec:rad}, we show that the hydrogen recombination lines are predominantly formed toward the outer edge of the cloud. The ramification is that that these lines are sensitive to the physical conditions of the gas, rather than the properties of the central engine of LRDs.
Simply put, the diagnostic power of hydrogen recombination lines for tracing the intrinsic ionizing spectrum, as clearly demonstrated in our toy model (\S\,\ref{subsec:toy_agn}), breaks down in the presence of dense gas.

Conversely, the helium recombination lines originate near the inner edge of the cloud, suggesting that these lines likely remain tightly coupled to the intrinsic ionizing spectrum, and thus could serve as a useful probe of the central ionizing engine.
Among the three helium recombination lines typically accessible with JWST/NIRSpec---\heiiuv, \heiiir, and \heiir---\heiir\ gets redshifted outside of the wavelength window observable with NIRSpec at $z>4$, while \heiiuv\ is heavily attenuated in the models and commonly appears in absorption since it is blue-ward of the Balmer limit.
\heiiir\ is thus left as the most promising probe.

To investigate the relationship between the recombination lines and the ionizing spectrum further, we examine the emergent line fluxes normalized by the number of ionizing photon for the corresponding atomic species from our \cloudy\ model grid.
If the emergent line flux traces the ionizing properties of the source, then this number should remain approximately constant. Such expectation is not met for \ha: the distributions of \ha\ flux per hydrogen-ionizing photon, $\qion$, have wide spreads spanning $\sim 2$~dex.
This result is consistent with our earlier conclusion that \ha\ (as well as other hydrogen recombination lines) is strongly influenced by reprocessing within the cloud, thereby disrupting the expected physical relationship between line flux and number of ionizing photons.

In contrast, the emergent \heii\ flux per \heii-ionizing photon, $f_{{\rm He}}/\qionhe$, exhibits considerably narrower distributions $\sim 0.5$~dex.
Therefore, although \heii\ emission can be more complex to interpret for reasons as explained in Section~\ref{sec:toy}, these results suggest that \heii\ emission, unlike hydrogen recombination lines, preserves at least some of the diagnostic power, even in the presence of dense gas.
Potential scenarios in which the gas becomes optically thick to \heii\ are further discussed in Section~\ref{sec:dis:heiithick}.

\subsection{Intrinsic Spectra of LRDs Hidden within a Dense Hydrogen Cloud\label{sec:dis:joint}}

The combination of weak \heii\ emission and a red rest-optical continuum makes it unlikely that the SEDs of LRDs resemble those of standard AGNs. While previous studies have hinted at this inconsistency (e.g., \citealt{Labbe2024:monster,Wang2024:brd,Sacchi2025}), in this paper we demonstrate that such a statement remains valid even under the dense gas hypothesis proposed by \citet{Inayoshi2025:dense}. This is particularly interesting given recent observational \citep{deGraaff2025:cliff,Ji2025,Naidu2025} as well as theoretical efforts \citep{Kido2025,Liu2025,Begelman2025} that attempt to explain LRDs as AGNs embedded in a dense gas envelope.

Moving beyond typical AGNs, our analysis suggests three possible interpretations for the nature of LRDs, driven by requiring consistency with the observed \heii\ EWs and the red optical continua. We organize the discussion of these scenarios to follow the sequence of results presented in Section~\ref{sec:res:joint}.

\subsubsection{A Previously Unseen, Optically Red AGN Accretion Disk\label{sec:alt_agn}}

An intrinsically optically red AGN can simultaneously satisfy the observed constraints on weak \heii\ and red rest-optical continua, without requiring fine-tuning of the ionizing spectral slope or invoking dust attenuation. However, the implied continuum would be nearly flat in $f_\lambda$ at optical wavelengths.
For context, such a red continuum corresponds approximately to a blackbody temperature of $T \sim 5000$~K, whereas the accretion disks of typical UV-luminous AGNs exhibit much hotter blackbody temperatures of $T \gtrsim 10^5$~K.

While a simple power-law does not accurately capture the optical continua of LRDs, particularly in cases where strong spectral breaks are present \citep{Hviding2025}, the few continuum slopes being explored are purely empirical and in need of better theoretical underpinnings. This exercise is intended to shed light on the required redness of the intrinsic optical slopes given the observational constraints. We do not attempt to link the redness of the continuum to a specific physical model in this work; rather, it serves as a useful illustration, motivated by recent developments in the theoretical front.

Although such a red AGN may appear exotic, several theoretical models have been proposed to account for this SED shape. Examples include super-Eddington accretion (e.g., \citealt{Lambrides2024,Kido2025,Liu2025,Zhang2025:mhd}), composite disks consisting of an inner standard disk and an outer gravitationally unstable accretion disk \citep{Zhang2025:disks}, and massive black hole binaries surrounded by circum-black hole and circum-binary disks, each with different effective temperatures \citep{Inayoshi2025:bbh}.

\subsubsection{Dust\label{sec:dust}}

At very high neutral hydrogen densities, the continuum gets completely reprocessed particularly at $\lambda < 4000$~\AA, but at redder wavelengths the continuum is merely reddened, which helps to relax the need for a large amount of dust. This is the premise behind the models explored in \citet{deGraaff2025:cliff,Naidu2025}.
While dust attenuation curves can vary and we have only explored a limited range of dust parameters, it is worth pointing out that it remains difficult for LRDs to exhibit an ionizing spectrum of a typical AGN, given the low \heii\ EW.
In other words, it would require a very soft ionizing spectrum and some dust to be consistent with observations.

As gas attenuation increases ($n({\rm{H}}) \gtrsim 10^{10.5}~{\rm cm^{-3}}$), the \heii\ EW also tends to rise, likely due to the interplay of two effects. First, the line flux increases with increasing hydrogen number density, $n({\rm{H}})$, because the ionization parameter scales as $U \propto Q({\rm{H}})/n({\rm{H}})$. At fixed $U$, a higher $n({\rm{H}})$ implies a larger number of ionizing photons, $Q({\rm{H}})$. Second, the continuum near \heii\ becomes increasingly absorbed as gas density increases. Although a higher $Q({\rm{H}})$ roughly sets the overall normalization of the emergent spectrum, meaning that this does not necessarily change the EW under normal, low-density conditions, the combined effect of enhanced line flux and suppressed continuum here possibly results in an elevated \heii\ EW.

The above leads to a tension: higher gas attenuation helps explaining the red continuum with a low $\Av$, but at the same time, it boosts the \heii\ EW toward a level that is inconsistent with observations. Thus, reconciling both the red continuum and weak \heii\ emission remains a challenge for models invoking standard AGN-like ionizing spectra.

\subsubsection{A Composite of AGN and Stars\label{sec:comp}}

With the Balmer break potentially produced by dense neutral hydrogen, one major tension in the stellar interpretation of LRDs is alleviated: the presence of a prominent Balmer break no longer necessarily implies evolved stellar populations in the early universe. Nonetheless, mixing in some young stars inside the dense hydrogen sphere around the AGN could also potentially explain the observed weak \heii\ emission.

Dense and massive assemblies of stars have been found at the centers of most galaxies (see e.g., \citealt{Neumayer2020} for a review). A notable example is the nuclear star cluster in the immediate vicinity of a massive black hole Sgr A* (e.g., \citealt{Bartko2010,Genzel2010}).
It has also been proposed the outer region of a standard steady-state accretion disk of a massive black hole could fragment into stars due to self-gravity \citep{Goodman2003}, suggesting that stars could reside within the same dense gaseous envelope. Such a nuclear star cluster scenario has been suggested as one possible explanation for the SED of ``The Cliff'' \citep{deGraaff2025:cliff}, while more recent work has shown that supermassive stars \citep{Nandal2025} or very dense stellar systems \citep{Bellovary2025} can reproduce some of the spectral signatures of LRDs.

Since stars typically produce a softer ionizing spectrum than AGNs, a composite SED with stellar contributions naturally results in lower \heii\ EWs that can be fully consistent with observations.
Although young SSPs also have blue rest-optical continua, the presence of dense neutral gas can attenuate the continuum. Therefore, a scenario in which stellar emission contributes substantially to the LRD SED offers a physically plausible explanation for both the weak \heii\ and the red continua without dust.

As a simple exercise, we adopt the synthetic SEDs of O-type and Wolf-Rayet stars \citep{Marcolino2017,Marcolino2024} available in the POLLUX database \citep{Palacios2010}, which are created with the 1-D radiative transfer code \texttt{CMFGEN} \citep{Busche2005}, as well as SEDs of stripped stars \citep{Goetberg2018}. We predict their emission-line EWs using \cue\ \citep{Li2024}, and find that these stellar spectra can produce similarly low \heii\ EWs as those in LRDs, though tend to over-predict the \ha\ EW.

Crucially, high hydrogen densities are highly conducive for star formation, and so stars can potentially form in the gas cloud.
More generally, the outer portions of most models of gas around the edge are gravitationally unstable, also more easily leading to star formation. 
Examinations of the relationship between the UV continua and Balmer emission lines in LRDs point in the same direction: both narrow and broad \ha\ components are tightly correlated with the UV continuum, with luminosity ratios consistent with those observed in young starburst galaxies \citep{Asada2026}.
The possibility of stellar contributions additionally provides a natural explanation for resolved UV morphologies observed in the bright LRDs from RUBIES \citep{Wang2024:brd,Baggen2024}, as well as in other studies \citep{Chen2025:host,Rinaldi2025}.

Given the common interpretation of the resolved fuzzy blue component seen in several LRDs as galaxy light, we estimate the stellar mass by scaling a young SSP to the UV part of the spectrum of \rd. The implied stellar mass is $\sim 10^{7.7}~\msun$. However, such SSPs are intrinsically very blue in the UV and therefore fail to reproduce the shape of the UV continuum of the LRD. Introducing a small amount of dust attenuation ($\Av = 0.5$) yields a good match to the UV continuum shape, with the implied stellar mass increasing to $\sim 10^{8.5}~\msun$. 
Interestingly, these values are not too far off from the stellar mass of $10^{7.7 \pm 0.3} \msun$ for the LRD sample of \citet{Matthee2024:cluster}, which are inferred based on clustering analysis. 
We can postulate a possible geometry where a thick gas cloud contains some newly formed stars. This is consistent with the scenario considered here that LRDs reside in low-mass, low-metallicity star-forming systems.
However, such configuration also depends on the accretion timescale, which is currently unknown. If the accretion timescale is much shorter than the timescale for star formation, then this scenario of a black hole with stars would no longer be viable.

\subsection{Limitations of the HeII Diagnostic\label{sec:dis:heiithick}}

It has been proposed that the broad lines observed in LRDs may result from electron scattering \citep{Begelman2025,Rusakov2025,Naidu2025}.
A key feature of this mechanism is the absorption near the systemic velocity, which has been observed in some LRDs (e.g., \citealt{Matthee2024}).
In addition, the broad Gaussian components of emission lines should exhibit similar widths, as the recombination or collisional excitation originates in similar regions within the broad-line region.

Recently, the differing line profiles for \ha, \hb, and Pa$\beta$ of a bright LRD were used as a counter argument for the electron scattering scenario \citep{Brazzini2025}. Whether this discrepancy holds across the broader LRD population remains an open question.
Nevertheless, because electron scattering is achromatic, the entire optical-UV SED would be optically thick to scattering. In such a case, even \heii\ emission may no longer serve as a clean tracer of the intrinsic spectrum.

However, we note that the optical depths of hydrogen and helium photons may be significantly lower far from the line center. As a result, the broad components of these lines could be less affected by the dense gas, potentially retaining diagnostic power for probing the central engine of LRDs.

An interesting implication of light scattering in such dense media is the potential for polarization. In a non-spherical geometry, emission lines from ionized gas would exhibit varying degrees of polarization depending on their formation sites; i.e., lines that undergo fewer electron scatterings would be less polarized. Electron scattering typically produces polarization at the level of $\sim 1$\% \citep{Poeckert1976}, and such effects have been observed in spectropolarimetric studies of hot massive stars (e.g., \citealt{Oudmaijer2005,Aret2016}).

Another possible mechanism is Stark broadening, which is the electric-field analog of the Zeeman effect and is typically considered in plasma physics \citep{Stark1914}. This is less discussed in the context of LRDs, and the resulting line profiles are yet to be investigated.

One aspect that is minimally discussed in this paper is the potential line-of-sight dependence. This may be particularly relevant given the detection of high-ionization lines in a subset of LRDs. For example, \citet{Labbe2024:monster} reported \nev\ emission in a highly luminous LRD, and \citet{Tang2025} found that narrow high-ionization lines are present in $\sim10$\% of LRDs. These results suggest that some LRDs likely have hard ionizing spectra. It is also plausible that LRDs more generally indeed have hard ionizing spectra, but the gas is so dense, possibly along certain sight-lines, that it is also optically thick to \heii. Alternatively, the \heii\ emission may be so broadened that it appears weak or effectively undetectable.

Beyond the limitations mentioned here, there are additional caveats in our current modeling approach. In particular, the approximation of a uniform-density gas slab is likely an over-simplification. Robustly capturing the relevant physics, such as radiation pressure, gravity, and electron scattering, requires more sophisticated modeling frameworks, akin to those typically employed in stellar atmosphere studies (e.g., \citealt{Hubeny1995,Hillier1998,Hamann2003,Lanz2003}).

\section{Conclusions\label{sec:concl}}

Thus far, little work has been done to directly constrain the intrinsic ionizing properties of LRDs. In this paper, we introduce a new approach based on helium recombination lines to do so. The main findings are summarized as follows.

First, recombination lines encode valuable information about the ionizing properties of a source, particularly in the case of AGNs. Unlike stellar populations, the ionizing spectrum of an AGN is thought to be well-approximated by a single power-law, allowing recombination lines to serve as more than mere ionizing photon counters. Using a power-law toy model, we demonstrate that a monotonic relationship exists between the spectral hardness and the strength of recombination lines. This relationship provides a conceptual framework for interpreting observed recombination lines as direct tracers of the intrinsic ionizing properties.

Applying this framework to a prototypical LRD, \rd\ \citep{Wang2024:ub}, reveals the first major puzzle: the object exhibits a high \ha\ EW, suggestive of an intrinsically very blue ionizing spectrum, harder than typical AGNs. Yet, the \heii\ emission is only marginally detected, pointing to a much softer spectrum this is more analogous to stellar models.

Taken at face value, the above suggests a peculiar ionizing spectrum---one capable of producing copious hydrogen-ionizing photons but few helium-ionizing photons; i.e., the implied ionizing spectrum must have a sharp turnover near the ionization potential of helium. The ionizing photon ratio, $Q_{\rm ion, H\textsc{i}}/Q_{\rm ion, He\textsc{ii}}$, typically has a value of $\sim 10$. However, this ratio increases to $\sim 300$ assuming the narrow component only \heii\ fit, and $\sim 50$ assuming the narrow and broad components fit.
The exceptionally low value of log~(\heii/\hb) of \rd\ ($\sim 10^{-2}$, $20 \times$ lower than the local AGN median) reinforces the conclusion that its ionizing spectrum is unusually soft, markedly different from that of standard AGNs.

Second, we assess how these diagnostics behave in the presence of dense neutral hydrogen gas, a hypothesis proposed in \citet{Inayoshi2025:dense} to explain the Balmer break of LRDs without a stellar origin. We find that the hydrogen recombination lines are heavily reprocessed within then dense medium, and therefore lose their ability to trace the intrinsic spectrum. Crucially, however, \heii\ emission retains its diagnostic power.

\heiiir, being generally the most accessible helium recombination line in JWST observations of LRD, thus emerges as a key tool for probing the intrinsic spectra of LRDs. 
Based on photoionization modeling with \cloudy, we rule out the possibility that the intrinsic SEDs of LRDs resemble those of typical AGNs, as none of the models can simultaneously reproduce the observed low \heii\ EW and the red rest-optical continuum shape. We explore several alternative scenarios that can account for the observed properties of LRDs, including intrinsically red AGNs, dust attenuation, and composite AGN-stellar populations. Each of these, or a combination thereof, provides a physically plausible explanation for the spectral features of LRDs.

Finally, a word of caution: \rd, and LRDs more generally, may instead have hard ionizing spectra.
However, the gas could be sufficiently dense, possibly along certain sight lines, to be optically thick to \heii; alternatively, the \heii\ emission may be so broadened that it becomes effectively undetectable in current data. Either effect could potentially explain the observed weak \heii\ emission.
Definitive detections and characterizations of high-ionization lines would point to an alternative scenario where LRDs do produce hard photons and these escape along sight-lines cleared of ultra-dense gas.

In summary, helium recombination lines, particularly \heiiir, can be used as a valuable probe of the intrinsic spectra of LRDs. Future works involving more sophisticated radiative transfer and kinematic modeling, combined with deeper medium- or high-resolution spectroscopy, will be crucial to further the progress in understanding LRDs, and subsequently, their place in the narrative of galaxy and black hole formation and evolution.

\section*{Acknowledgments}

B.W. thanks Michael Eracleous for valuable discussions.
B.W. and J.L. acknowledge support from JWST-GO-04233.009.
B.W. also acknowledges support provided by NASA through Hubble Fellowship grant HST-HF2-51592.001 awarded by the Space Telescope Science Institute, which is operated by the Association of Universities for Research in Astronomy, In., for NASA, under the contract NAS 5-26555.
K.I. acknowledges support from the National Natural Science Foundation of China (12573015, W2532003), the Beijing Natural Science Foundation (IS25003), and the China Manned Space Program (CMS-CSST-2025-A09).
R.E.H. acknowledges support by the German Aerospace Center (DLR) and the Federal Ministry for Economic Affairs and Energy (BMWi) through program 50OR2403 `RUBIES'.

This work is based on observations made with the NASA/ESA/CSA James Webb Space Telescope. The data were obtained from the Mikulski Archive for Space Telescopes at the Space Telescope Science Institute, which is operated by the Association of Universities for Research in Astronomy, Inc., under NASA contract NAS 5-03127 for JWST. These observations are associated with program \# 1433, 2561, 4106, 4233, 5224, 6585.
The specific observations analyzed can be accessed via 
\dataset[10.17909/9hpc-nc45]{https://doi.org/10.17909/9hpc-nc45}.
Computations for this research were performed on the Pennsylvania State University's Institute for Computational and Data Sciences' Roar supercomputer;
and on computational resources managed and supported by Princeton Research Computing, a consortium of groups including the Princeton Institute for Computational Science and Engineering (PICSciE) and Research Computing at Princeton University.
Some of the stellar spectra are retrieved from the POLLUX database (pollux.oreme.org) operated at LUPM (Universit\'{e} de Montpellier - CNRS, France) with the support of the PNPS and INSU.
This publication made use of the NASA Astrophysical Data System for bibliographic information.

\facilities{JWST (NIRSpec)}
\software{
Astropy \citep{2013A&A...558A..33A,2018AJ....156..123A,2022ApJ...935..167A},
Cloudy \citep{Ferland2017},
FSPS \citep{Conroy2010},
Matplotlib \citep{2007CSE.....9...90H},
NumPy \citep{2020Natur.585..357H}
}

\section*{Appendix}
\counterwithin{figure}{section}
\counterwithin{table}{section}
\renewcommand{\thesection}{\Alph{section}}
\setcounter{section}{0}

\section{Samples of LRDs\label{app:sample}}

Motivated by the diversity observed in LRDs, we include additional examples from the literature to check the generality of our findings.
All selected objects have been studied in detail in the past. This is an intentional choice, as the various studies can provide valuable context for comparison and interpretation.

Specifically, the sources outside the RUBIES program include:
MoM-BH*-1, which has one of the strongest Balmer breaks detected to date \citep{Naidu2025};
A2744-QSO1, locating in a strongly lensed area and thus has a well constrained size of merely $\sim 30$~pc \citep{Furtak2024};
A2744-45924, a remarkably luminous LRD showing signs of an \feii\ forrest in the rest UV \citep{Labbe2024:monster};
J0647\_1045 \citep{Killi2024}, which exhibits a Balmer-jump-like feature;
and COS-66964, detected with rest-UV emission lines \citep{Akins2025:uv}.

\begin{figure*} 
\gridline{
  \fig{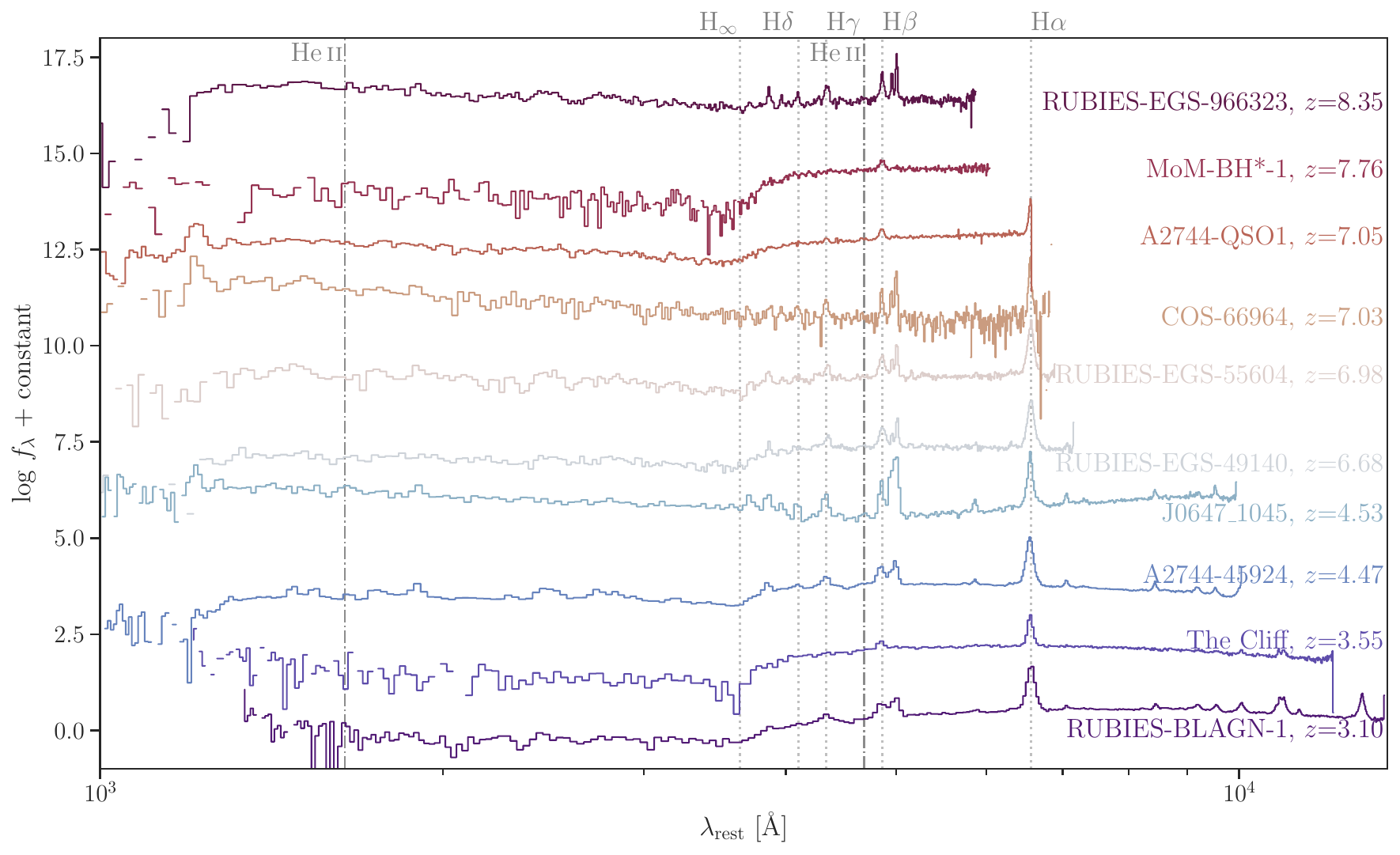}{0.95\textwidth}{}
}
\caption{JWST/Prism spectra of some of the LRDs that have been studied in detail in various studies.}
\label{fig:app:sample}
\end{figure*}

\begin{figure*} 
\gridline{
  \fig{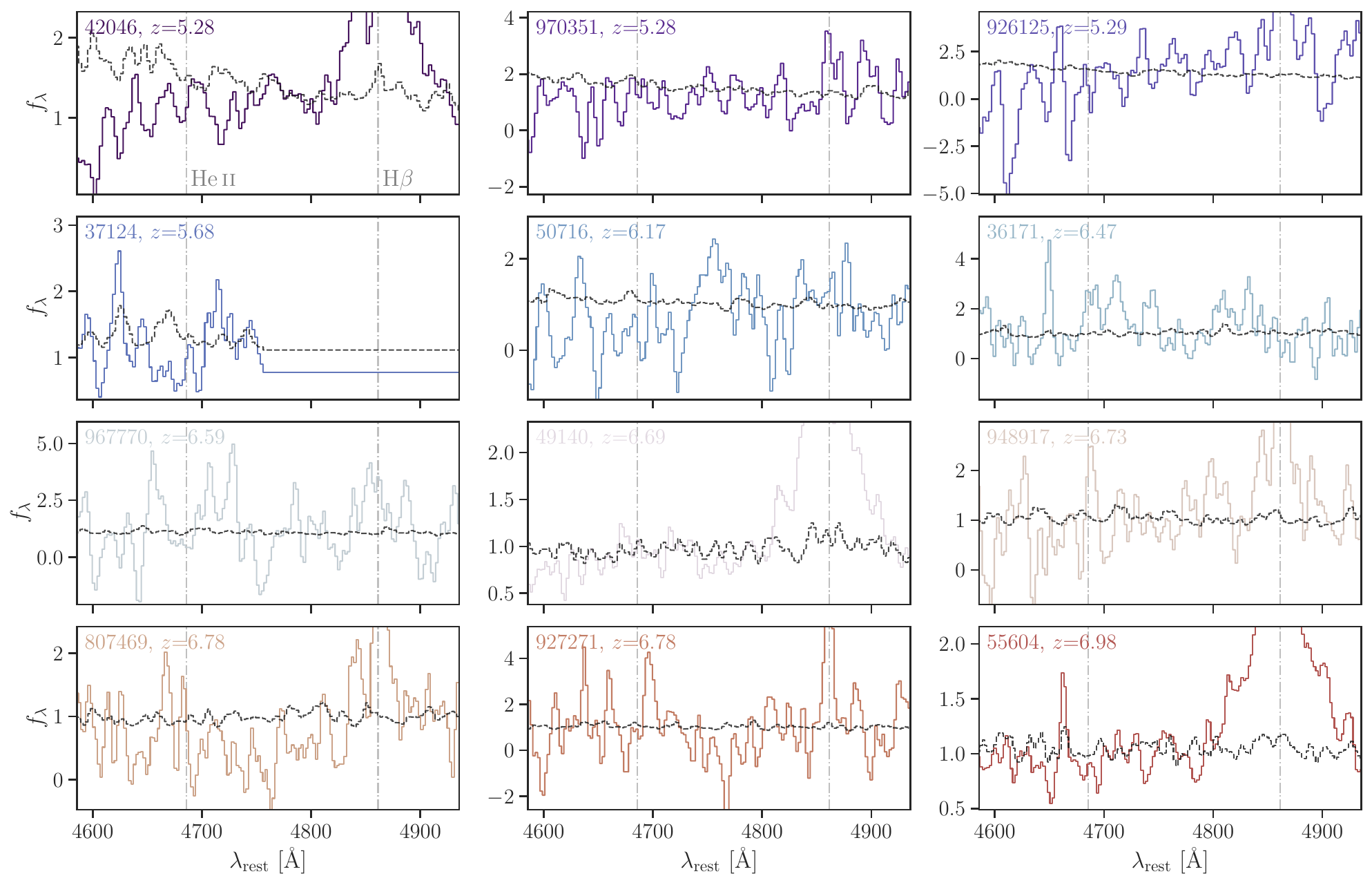}{0.95\textwidth}{}
}
\caption{Medium-resolution G395M spectra for some of the bright LRDs in the RUBIES sample are shown in colored lines \citep{Hviding2025}, whereas the noises are over-plotted as black dashed lines.
These spectra demonstrate that the low \heii\ EWs found in LRDs in general are not an effect of the low resolution of Prism data.
}
\label{fig:app:rubies}
\end{figure*}

Information on each source is listed in Table~\ref{tab:app:sample}, and their Prism spectra are shown in Figure~\ref{fig:app:sample}.
In all cases, the \heii\ appears weak or not detected.
The G395M spectra further demonstrate that the low \heii\ EWs found in LRDs in general are not an effect of the low resolution of Prism data.
While we refer readers to \citet{Hviding2025} for a full description of the RUBIES LRD sample, here we highlight the \heii\ region in the G395M spectra for a subset in Figure~\ref{fig:app:rubies}.

\begin{deluxetable*}{llll}
\tablecaption{Notable LRDs\label{tab:app:sample}}
\tablehead{
\colhead{ID} & \colhead{$\zspec$} & \colhead{JWST PID} & \colhead{Reference for spectral modeling}
}
\startdata
RUBIES-EGS-49140 & 6.68351 & GO-4233 & \citet{Wang2024:ub} \\
RUBIES-EGS-55604\tablenotemark{\scriptsize{a}} & 6.98173 & GO-4233 &\citet{Wang2024:ub} \\
RUBIES-EGS-966323\tablenotemark{\scriptsize{a}} & 8.3539 & GO-4233 & \citet{Wang2024:ub} \\
RUBIES-BLAGN-1 (RUBIES-UDS-40579) & 3.1034 & GO-4233 & \citet{Wang2024:brd} \\
The Cliff (RUBIES-UDS-154183) & 3.546 &  GO-4233 & \citet{deGraaff2025:cliff} \\
A2744-QSO1\tablenotemark{\scriptsize{b}} & 7.0451 & GO-2561 & \citet{Ma2025,Ji2025} \\
A2744-45924 & 4.47 &  GO-2561 & \citet{Labbe2024:monster} \\
J0647\_1045 & 4.53 & GO-1433 &\citet{Killi2024} \\
MoM-BH*-1 (MoM-UDS-150135) & 7.7569 &  GO-5224 & \citet{Naidu2025} \\
COS-66964 & 7.0348 & DD-6585 &\citet{Akins2025:uv} \\ \enddata
\tablenotetext{a}{Photometric modeling presented in \citet{Labbe2023:ub}.}
\tablenotetext{b}{JWST/Prism spectra presented in \citet{Furtak2024}.}
\end{deluxetable*}

\section{Uncertainties in HeII EW\label{app:linefit}}

\begin{figure*} 
\gridline{
  \fig{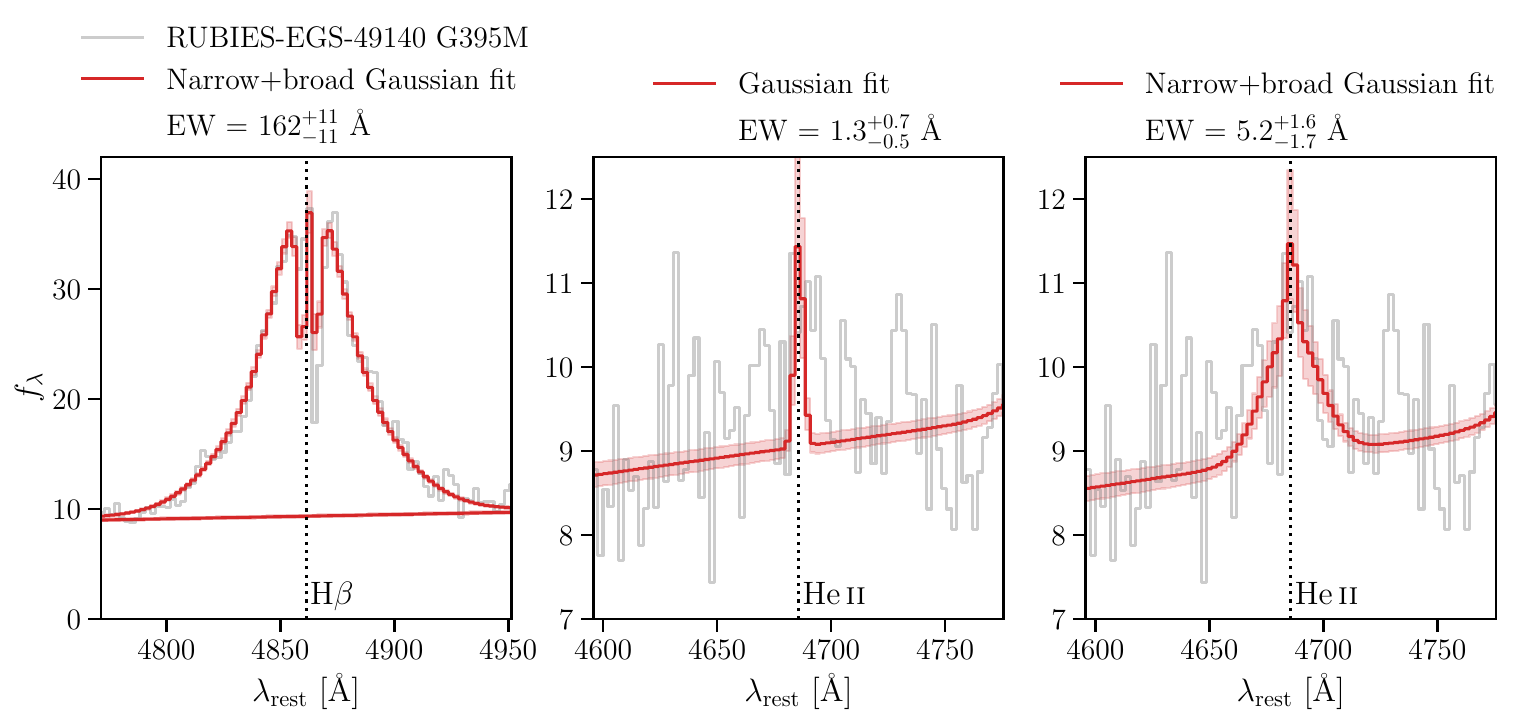}{0.95\textwidth}{}
}
\caption{Line profile fits for \rd, highlighting the weakness in \heii\ emission. In each panel, the observed G395M spectrum is shown in gray, and the best-fit model is over-plotted in red. All lines are fit simultaneously. The narrow components are tied to share the same velocity dispersion, and likewise for the broad components.
The second panel shows the fit to the \heii\ line using only a narrow Gaussian component, while the third panel includes both a narrow and a broad component.
These two fits serve to encapsulate the systematic uncertainties in the \heii\ measurements arising from the unknown kinematics.
}
\label{fig:app:heii}
\end{figure*}

Measuring \heii\ is challenging due to potential line blending (e.g., with He\,\textsc{i}\,4676 or [Ar\,\textsc{iv}]\,4711), the ambiguity in the local continuum, and the uncertain kinematics.
The former two issues are illustrated in Figure~\ref{fig:data}, where we show the Prism and medium-resolution spectra of \rd, along with the model fit. 
In this work, we define the local continuum to be $\sim 4600 - 4750$~\AA. However, there appears to be a bump over the wider wavelength range of $\sim 4500 - 4800$~\AA, resembling a Wolf-Rayet feature observed in starburst galaxies (e.g., \citealt{Lopez-Sanchez2010}), and may signal substantial outflow.
Certainly if one defines the local continuum by selecting the lowest-flux region around 4550~\AA, the EW would be higher, reaching $>10$~\AA.

Figure~\ref{fig:app:heii} further demonstrates the uncertainty due to unknown kinematics. Assuming that the \heii\ emission only has a narrow component leads to an EW of $1.3 \pm 0.6$~\AA, whereas the addition of a broad component increases the EW to $5.2 \pm 1.7$~\AA.

Moreover, uncertainty can also arise from possible contamination by \feii, observed in the LRD presented in \citet{Labbe2024:monster}.
The \feii\ complex is a blend of multiple lines emitted under different conditions and at varying distances from the black hole, and the amount of line flux varies widely across AGN spectra. 
Crucially, \feii\ multiplet emission may masquerade as a pseudo-continuum \citep{Popovic2019}, biasing the \heii\ EW in either directions.

In the case of \rd, there is no clear evidence for emission from the consistent \feii\ groups overlapping the \heii\ region. However, a more flexible \feii\ model including the so-called inconsistent lines, i.e., the lines whose relative intensities could not be described with standard photoionization models \citep{Kovacevic-Dojcinovic2025}, can produce a dip at $\sim4550$~\AA. This feature may arise, for instance, in the presence of a strong \feii\ I Zw 1 group combined with a weak F-group \citep{Vestergaard2001}.

With upcoming deep, medium-resolution spectra, it will be possible to revisit our measurements of \heii. 
Future modeling efforts that incorporate detailed \feii\ templates may also help mitigate these uncertainties and allow for a more accurate determination of the \heii\ EW.


\section{Turbulent Velocity\label{app:vturb}}

\begin{figure*} 
\gridline{
  \fig{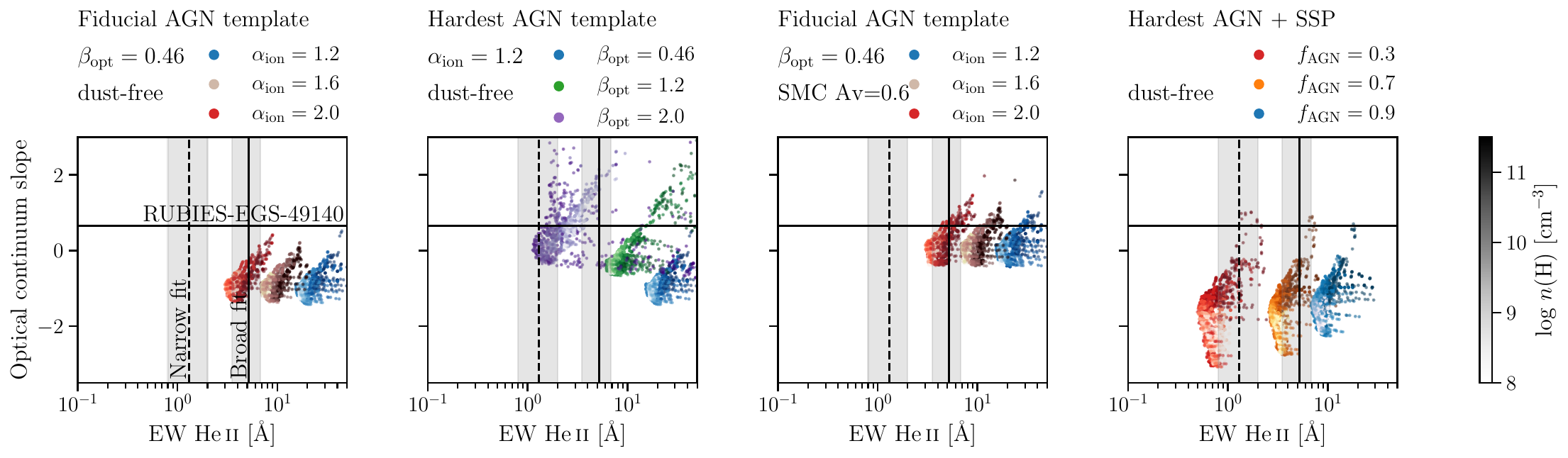}{0.95\textwidth}{}
}
\caption{Same as Figure~\ref{fig:slope}, but for $\vturb=500~\kms$. Main results of this paper remain unaffected with a higher turbulent velocity.}
\label{fig:app:heii_slp}
\end{figure*}

A turbulent velocity, $\vturb$, of $120~\kms$ is adopted in the main text, consistent with the best-fit value reported by \citet{Ji2025}. \citet{Naidu2025} find a higher value of $\vturb \sim 500~\kms$.
To explore the impact of different $\vturb$, we show model predictions corresponding to the more turbulent case in Figure~\ref{fig:app:heii_slp}.
Our main results remain unaffected.

\bibliography{lrd_qion_wang.bib}
\bibliographystyle{aasjournal}

\end{CJK*}
\end{document}